\documentclass[a4paper,11pt]{article}
\pdfoutput=1
\usepackage{jcappub}

\usepackage{amsthm}
\usepackage{graphicx}
\usepackage{xcolor}
\usepackage{physics}
\usepackage{braket}
\usepackage{amsmath}
\usepackage{amssymb}
\usepackage{enumerate}
\usepackage{mathtools}
\usepackage{etoolbox}
\usepackage{algorithmic}
\usepackage{algorithm}
\usepackage{bbm}
\usepackage{here}
\usepackage{comment}
\usepackage{bm}
\usepackage{acronym}
\usepackage{siunitx}
\usepackage{ragged2e}
\usepackage{booktabs}

\newcounter{algo}

\newcommand{\argmin}{\mathop{\rm argmin}\limits}

\newcommand{\beae}[1]{\begin{equation}\begin{aligned} #1 \end{aligned}\end{equation}}
\newcommand{\bege}[1]{\begin{equation}\begin{gathered} #1 \end{gathered}\end{equation}}
\newcommand{\bae}[1]{\begin{align} #1 \end{align}}
\newcommand{\bce}[1]{\begin{cases} #1 \end{cases}}

\newcommand{\bme}[1]{\begin{multline} #1 \end{multline}}
\newcommand{\bmte}[1]{\begin{multlined}[t] #1 \end{multlined}}
\newcommand{\bmbe}[1]{\begin{multlined}[b] #1 \end{multlined}}

\newcommand{\relmiddle}[1]{\mathrel{}\middle#1\mathrel{}}

\newcommand{\doublevev}[1]{\expval{\!\!\expval{#1}\!\!}}

\definecolor{MONZA}{HTML}{CF000F}
\definecolor{DARKBLUE}{HTML}{00008b}
\definecolor{DARKMAGENTA}{HTML}{8b008b}

\DeclareMathOperator{\Var}{Var}
\DeclareMathOperator{\Cov}{Cov}
\DeclareMathOperator{\Mean}{\mathbb{E}}
\newcommand{\bk}{\mathrm{bk}}
\newcommand{\PS}{\mathrm{PS}}
\newcommand{\Mpl}{M_\mathrm{Pl}}
\newcommand{\uend}{\mathrm{end}}
\newcommand{\ns}{n_\mathrm{s}}
\newcommand{\well}{\mathrm{well}}
\newcommand{\ini}{\mathrm{ini}}

\newcommand{\uc}{\mathrm{c}}
\newcommand{\sfH}{\mathsf{H}}
\newcommand{\bfk}{\mathbf{k}}
\newcommand{\calN}{\mathcal{N}}
\newcommand{\calP}{\mathcal{P}}
\newcommand{\uw}{\mathrm{w}}
\newcommand{\bfX}{\mathbf{X}}
\newcommand{\calX}{\mathcal{X}}

\makeatletter
\newcommand{\preprintnumber}[1]{%
  \gdef\@preprintnumber{#1}}
\def\@preprintnumber{}
\def\printpreprint{%
  \begin{flushright}
    \small \@preprintnumber
  \end{flushright}
}
\makeatother

\preprintnumber{RUP-25-14}

\abstract{
The stochastic-$\delta \calN$ formalism is widely used to study inflation models in which the quantum diffusion of inflatons dominates the background dynamics, leading to interesting phenomena such as the production of primordial black holes.
Among numerical approaches to calculate the curvature perturbation spectrum $\calP_\zeta(k)$ in this formalism, the Monte Carlo simulation-based approach has been proposed as a promising choice, especially in multifield cases.
In this approach, we generate many paths of inflatons from the initial points to the end of inflation, obtain statistics of $\delta N$ from the paths, and then estimate $\calP_\zeta(k)$.
However, this method involves a nested Monte Carlo simulation, which requires generating many branch paths from each trunk path at the point corresponding to the scale $k$ of interest, resulting in a high computational cost.
In this paper, we propose a new Monte Carlo-based approach that utilizes least squares fitting, introducing two novel features for reducing computational cost.
First, we devise a simple estimator of a key statistic $\expval{\delta \mathcal{N}_{\mathbf{X}}^2}$, the variance of $\delta \calN$ conditioned on the branching point, to avoid nesting path generation.
Second, via least squares fitting of a parametric function to the sampled values of the estimator, we obtain not just an estimate of $\calP_\zeta(k)$ for a single value of $k$ but an approximating function of $\calP_\zeta(k)$ over a range of $k$ of interest.
We also conduct numerical demonstrations for concrete inflation models, which show the usefulness of our method.
}

\begin{document}

\title{\boldmath Calculating the power spectrum in stochastic inflation by Monte Carlo simulation and least squares curve fitting}

\author[a]{Koichi Miyamoto}
\author[b,c,d]{and Yuichiro Tada}

\affiliation[a]{Center for Quantum Information and Quantum Biology, The University of Osaka, Toyonaka 560-0043, Japan}
\affiliation[b]{Department of Physics, Rikkyo University, Toshima, Tokyo 171-8501, Japan}
\affiliation[c]{Institute for Advanced Research, Nagoya University,
Furo-cho, Chikusa-ku, 
Nagoya 464-8601, Japan}
\affiliation[d]{Department of Physics, Nagoya University,
Furo-cho, Chikusa-ku,
Nagoya 464-8602, Japan}

\emailAdd{miyamoto.kouichi.qiqb@osaka-u.ac.jp}
\emailAdd{yuichiro.tada@rikkyo.ac.jp}

\printpreprint

\date{\today}

\maketitle

\acrodef{PDF}{probability density function}
\acrodef{SDE}{stochastic differential equation}
\acrodef{EOI}{end of inflation}
\acrodef{MS}{Mukhanov--Sasaki}
\acrodef{EM}{Euler--Maruyama}
\acrodef{AV20}{Ando and Vennin}
\acrodef{FKTT}{Fujita, Kawasaki, Tada, and Takesako}
\acrodef{AV24}{Animali and Vennin}
\acrodef{TY}{Tada and Yamada}
\acrodef{USR}{ultra-slow-roll}
\acrodef{PDE}{partial differential equation}
\acrodef{STOLAS}{STOchastic LAttice Simulation}
\acrodef{FOREST}{FOrtran Recursive Exploration of Stochastic Trees}

\section{Introduction}

Inflation, the accelerated expansion of space in the early universe driven by certain scalar fields called inflatons, is a standard paradigm in cosmology.
One of its virtues, along with solving the horizon and flatness problems, is that it produces primordial density perturbations, which result in structures in today's universe, such as galaxies and clusters, through inflatons' quantum fluctuations.

Among various approaches to analyzing the inflationary primordial perturbations, the $\delta N$ formalism~\cite{Starobinsky:1985ibc,Salopek:1990jq,Sasaki:1995aw,Wands:2000dp,Lyth:2004gb}, 
which relates fluctuations of the e-fold number $N$ during inflation in different spatial patches 
to primordial perturbations, is one of the standard ones.
In particular, it is very helpful in the context of \emph{stochastic inflation}.
Stochastic inflation (see Refs.~\cite{Starobinsky:1982ee,Starobinsky:1986fx,Nambu:1987ef,Nambu:1988je,Kandrup:1988sc,Nakao:1988yi,Nambu:1989uf,Mollerach:1990zf,Salopek:1990re,Linde:1993xx,Starobinsky:1994bd} for the first works and also Ref.~\cite{Cruces:2022imf} for a recent review) is a probabilistic formalism for long-wavelength modes of fields.
It treats $\boldsymbol{\mathcal{X}}=(\boldsymbol{\varphi},\boldsymbol{\varpi})$, inflatons and their conjugate momenta coarse-grained on a superHubble scale, and describes their stochastic time-evolution driven by quantum fluctuations with the Langevin equation.
The distributions of the e-fold number and the large-scale primordial perturbation are accordingly determined.

This \emph{stochastic-$\delta \calN$ formalism}~\cite{Fujita:2013cna,Fujita:2014tja,Vennin:2015hra,Ando:2020fjm,Animali:2024jiz} has been used to study many interesting cosmological scenarios.
In particular, it takes advantage when the ordinary linear perturbation analysis, which assumes that the background dynamics dominate the fluctuations, does not work well, e.g., in the diffusion-dominated scenario, where inflatons traverse a very flat region in their potential, and thus their stochastic movement dominates over the deterministic slow-roll.
Such a scenario has attracted attention because amplitudes of the primordial curvature perturbation $\zeta$ of the corresponding wavelengths can be amplified, which leads to various cosmological phenomena such as the production of primordial black holes, a candidate for dark matter.
In fact, in analytical and numerical approaches, previous studies~\cite{Kawasaki:2015ppx,Assadullahi:2016gkk,Vennin:2016wnk,Pattison:2017mbe,Ezquiaga:2018gbw,Noorbala:2018zlv,Firouzjahi:2018vet,Noorbala:2019kdd,Kitajima:2019ibn,Prokopec:2019srf,Ezquiaga:2019ftu,Firouzjahi:2020jrj,De:2020hdo,Figueroa:2020jkf,Pattison:2021oen,Figueroa:2021zah,Tada:2021zzj,Ezquiaga:2022qpw,Ahmadi:2022lsm,Nassiri-Rad:2022azj,Animali:2022otk,Tomberg:2022mkt,Gow:2022jfb,Rigopoulos:2022gso,Briaud:2023eae,Asadi:2023flu,Tomberg:2023kli,Tada:2023fvd,Tokeshi:2023swe,Raatikainen:2023bzk,Miyamoto:2024hin,Tokeshi:2024kuv,Kuroda:2025coa,Takahashi:2025hqt} have performed stochastic inflation-based analyses in aforementioned scenarios, aiming to calculate quantities related to cosmological phenomena and observations, such as the power spectrum $\mathcal{P}_\zeta(k)$, which represents the amplitude of primordial perturbations of scale $k$.

However, extending such analyses to general cases, especially multifield models, is practically challenging, although there are well-motivated multifield diffusion-dominated scenarios such as mild waterfall transition in hybrid inflation~\cite{Fujita:2014tja,Kawasaki:2015ppx,Tada:2023pue,Tada:2023fvd,Tada:2024ckk}.
Since analytical approaches are basically applicable only to single-field cases, we often resort to numerical ones.
A numerical approach that is seemingly appealing is the Monte Carlo simulation-based method proposed in Ref.~\cite{Fujita:2014tja}, where we generate many paths of $\boldsymbol{\mathcal{X}}$'s time-evolution from an initial point to the \ac{EOI} and use them to calculate statistical quantities such as $\mathcal{P}_\zeta(k)$.
Since Monte Carlo simulation is often used to analyze high-dimensional random processes, avoiding the blow-up of the computational cost, this method seems a good choice in multifield cases.
However, this method suffers from high computational cost, whether the inflation is single or multiple, because it requires \emph{nested Monte Carlo simulation}, whose procedure is as follows.
We first generate many paths.
On each of them, we take points corresponding to $k$'s for which we want to calculate $\mathcal{P}_\zeta(k)$.
Then, from these points, we generate many branch paths.
The total number of paths including branches is of order $O(n_{\mathrm{path},1}n_{\mathrm{path},2}n_{\rm PS})$, where $n_{\mathrm{path},1}$ is the number of trunk paths, $n_{\mathrm{path},2}$ is the branch number per branching point, and $n_{\rm PS}$ is the number of $k$'s.

In this paper, we propose a novel Monte Carlo-based method to calculate $\mathcal{P}_\zeta(k)$ built upon the formula given by \ac{AV20} in Ref.~\cite{Ando:2020fjm}. 
Although we leave the detail to Secs.~\ref{sec:prel} and \ref{sec:method}, according to the formula, we can find $\mathcal{P}_\zeta(k)$ via calculating $\expval{\Var\left[\calN_{\bm{\calX}_{N_\bk}}\relmiddle{|}\bm{\calX}_{N_\bk}\right]}$, the variance of the e-fold number of paths starting from $\boldsymbol{\mathcal{X}}_{N_{\rm bk}}$, $\boldsymbol{\mathcal{X}}$ at the backward e-fold $N_{\rm bk}$, averaged with respect to $\boldsymbol{\mathcal{X}}_{N_{\rm bk}}$.
Here, the backward e-fold means the e-fold counted from the \ac{EOI} along the path, and the value of $N_{\rm bk}$ has a one-to-one correspondence with $k$, the wavenumber of interest, in our algorithm.
Since this is a kind of nested expectation, we are seemingly required to run a nested Monte Carlo simulation.
Fortunately, as proposed in Sec.~\ref{sec:method}, we can take an estimator of $\Var\left[\calN_{\bm{\calX}_{N_\bk}}\relmiddle{|}\bm{\calX}_{N_\bk}\right]$ by $\frac{1}{2}\left(\mathcal{N}_{\boldsymbol{\mathcal{X}}_{N_{\rm bk}}}^{(1)}-\mathcal{N}_{\boldsymbol{\mathcal{X}}_{N_{\rm bk}}}^{(2)}\right)^2$, where $\mathcal{N}_{\boldsymbol{\mathcal{X}}_{N_{\rm bk}}}^{(1,2)}$ are the e-fold numbers of two paths that branch at backward e-fold $N_{\rm bk}$.
We can calculate this by generating just two branches at each branching point, which means that the number of paths immediately reduces.
We further make an improvement on the computational cost with respect to $n_{\rm PS}$, leveraging \emph{least squares curve fitting}.
On each trunk path, we generate two branches not from prefixed values of $N_{\rm bk}$ but from \emph{one} randomly chosen value of $N_{\rm bk}$.
We then obtain a set of $\left(N_{\rm bk}, \frac{1}{2}\left(\mathcal{N}_{\boldsymbol{\mathcal{X}}_{N_{\rm bk}}}^{(1)}-\mathcal{N}_{\boldsymbol{\mathcal{X}}_{N_{\rm bk}}}^{(2)}\right)^2\right)$, and perform least squares fitting to this sample set, using some parametric function.
This yields not the values of the power spectrum at a finite number of $k$ but its \emph{approximate function} over a range of scales.
The number of paths generated in our method is just $O(n_{\rm path,1})$, which means a large reduction of the computational cost.

After presenting our method in full detail in Sec.~\ref{sec:method}, in Sec.~\ref{sec:result}, we perform its numerical demonstrations in four test cases.
The first three are single-field models, for which the precise power spectrum is obtained by other methods, and thus, we can make a comparison to check our method. 
The last one is hybrid inflation, one of the multifield models, in which our method is particularly desired. 
The results demonstrate that our method works in the tested cases to some extent.

\section{Preliminary \label{sec:prel}}

\subsection{Stochastic inflation}

We consider the general relativistic system of $\boldsymbol{\phi}=(\phi_1, \ldots, \phi_d)$, $d$ canonical scalar fields called inflatons,\footnote{We supposed the canonical kinetic term for the inflatons for simplicity. The stochastic formalism can be generalized to curved-target-space models; see Ref.~\cite{Pinol:2020cdp}.} described by the following action
\begin{align}
    S = \int\dd[4]{x}\sqrt{-g}\left[\frac{1}{2}R-\frac{1}{2}\sum_{i=1}^dg^{\mu\nu}\partial_\mu\phi_i\partial_\nu\phi_i-V(\boldsymbol{\phi})\right],
\end{align}
where $R$ is the Ricci scalar associated with the spacetime metric $g_{\mu\nu}$, $V(\boldsymbol{\phi})$ is the potential of $\boldsymbol{\phi}$, and we adopt the Planck unit where $c=\hbar=\Mpl=1$ ($\Mpl$ is the reduced Planck mass).
With the use of the e-folds $N\coloneqq\ln a$ defined by the scale factor $a$ as the time variable and the flat slicing for the equal-time hypersurface,\footnote{See, e.g., Refs.~\cite{Finelli:2008zg,Finelli:2010sh,Pattison:2019hef} for discussions about e-folding number as the time variable in the stochastic formalism.} we obtain the following Langevin equation that describes the time evolution of $\boldsymbol{\mathcal{X}}\coloneqq(\boldsymbol{\varphi},\boldsymbol{\varpi})$, the superHubble parts of $\boldsymbol{\phi}$ and $\boldsymbol{\phi}$'s conjugate momenta $\boldsymbol{\pi}\coloneqq(\pi_1,\ldots,\pi_n)$: for each $i=1,\ldots,d$,
\beae{
    &\dv{\varphi_i(N)}{N}=\frac{\varpi_i (N)}{H(\boldsymbol{\varphi}(N), \boldsymbol{\varpi}(N))} + \xi_{\phi_i}(N), \\
    &\dv{\varpi_i(N)}{N}=-3\varpi_i(N)-\frac{\partial_{\phi_i}V(\boldsymbol{\varphi}(N))}{H(\boldsymbol{\varphi}(N), \boldsymbol{\varpi}(N))}+\xi_{\pi_i}(N).
    \label{eq:SDE}
}
$H\coloneqq\dot{a}/a$ is the Hubble parameter given via the Friedmann equation
\begin{equation}
    3H^2(\boldsymbol{\phi}, \boldsymbol{\pi}) = \frac{1}{2} \sum_{i=1}^d\pi_i^2(N) + V(\boldsymbol{\phi}).
\end{equation}
$\boldsymbol{\varphi}=(\varphi_1,\ldots,\varphi_d)$ and $\boldsymbol{\varpi}=(\varpi_1,\ldots,\varpi_d)$ are defined by coarse-graining:
\beae{
    &\varphi_i(N,\mathbf{x})\coloneqq\int\frac{\dd[3]{k}}{(2\pi)^3} e^{i \mathbf{k}\cdot\mathbf{x}} \tilde{\phi}_i(N,\mathbf{k})\Theta(k_\sigma(N)-k), \\
    &\varpi_i(N,\mathbf{x})\coloneqq\int\frac{\dd[3]{k}}{(2\pi)^3} e^{i \mathbf{k}\cdot\mathbf{x}} \tilde{\pi}_i(N,\mathbf{k})\Theta(k_\sigma(N)-k).
}
Here, $\tilde{\phi}_i(N,\mathbf{k}) \coloneqq \int\dd[3]{x}e^{-i \mathbf{k}\cdot\mathbf{x}} \phi_i(N,\mathbf{x})$ (resp. $\tilde{\pi}_i(N,\mathbf{k}) \coloneqq \int\dd[3]{x}e^{-i \mathbf{k}\cdot\mathbf{x}} \pi_i(N,\mathbf{x})$) is the Fourier mode of $\phi_i$ (resp. $\pi_i$) with wavenumber vector $\mathbf{k}$.
$\Theta$ is the Heaviside function that takes $\Theta(z)=1$ for $z>0$ and 0 for $z<0$.\footnote{The consistent definition of $\Theta(z)$ at $z=0$ requires a detailed discussion of the discretization of the path
integral. See Refs.~\cite{Tokuda:2017fdh,Tokuda:2018eqs}.}
$k_\sigma(N)$ is defined by $k_\sigma(N)\coloneqq \sigma a(N) \mathsf{H}$ with the dimensionless and dimensionful parameters $\sigma$ and $\mathsf{H}$, which are set so that $k_\sigma(N) \ll a(N) H(\boldsymbol{\varphi}(N), \boldsymbol{\varpi}(N))$ holds for typical sample paths in the range of $N$ under consideration.\footnote{The coarse-graining scale $k_\sigma$ is often defined by $\sigma a(N)H(\bm{\varphi}(N),\bm{\varpi}(N))$ in the literature, but it requires a circular definition of $\bm{\varphi}$ and $\bm{\varpi}$. In this work, we rather use a certain constant $\sfH$ as a model parameter. It is less physical but can avoid the circular definition, and the original algorithm for the power spectrum proposed in Ref.~\cite{Ando:2020fjm} and adopted in this work anyway needs to suppose that the Hubble parameter is almost constant. See also footnote~3 of Ref.~\cite{Mizuguchi:2024kbl}.}
In this paper, we set $\sigma=0.1$ and $\mathsf{H}=H(\boldsymbol{\phi}_{\rm ini}, \boldsymbol{\pi}_{\rm ini})$, assuming that $(\boldsymbol{\phi},\boldsymbol{\pi})$ takes a globally equal initial value $(\boldsymbol{\phi}_{\rm ini},\boldsymbol{\pi}_{\rm ini})$. 
$\xi_X$ ($X=\phi_i$ or $\pi_i$) are the white Gaussian noises with zero means and covariances
\begin{align}\label{eq: noise correlation}
    \expval{\xi_X(N)\xi_Y(N')}=\calP_{XY}(k_\sigma(N))\delta(N-N'),
\end{align}
for $X,Y=\phi_i,\pi_i$, where the power spectrum $\mathcal{P}_{XY}$ defined by
\bae{\label{eq: calP}
    \expval{\hat{X}_\bfk\hat{Y}_{\bfk'}}=(2\pi)^3\delta^{(3)}(\bfk+\bfk')\frac{2\pi^2}{k^3}\calP_{XY}(k),
}
with the corresponding quantum operators $\hat{X}_\bfk$ and $\hat{Y}_\bfk$ in Fourier space. While $\expval{\cdot}$ represents the quantum average in Eq.~\eqref{eq: calP}, it denotes the ensemble average in Eq.~\eqref{eq: noise correlation} and hereafter.
Note that although $\boldsymbol{\mathcal{X}}$ (and also $\boldsymbol{\xi}$) takes different values at different spatial points $\mathbf{x}$, we have omitted $\mathbf{x}$ in Eq.~\eqref{eq:SDE}.
This is because we will hereafter solve Eq.~\eqref{eq:SDE} without considering $\mathbf{x}$ dependence, and regard each sample path as representing a realization of the time-evolving $\boldsymbol{\mathcal{X}}(N)$ at one spatial point.

When we generate a sample path of $\boldsymbol{\mathcal{X}}$, we need to discretize Eq.~\eqref{eq:SDE} in time by some method.
In this paper, we adopt the \ac{EM} method~\cite{Maruyama1955} given as 
\beae{
    &\bmte{\varphi_i(N+\Delta N) =\varphi_i(N) + \frac{\varpi_i(N)}{H(\boldsymbol{\varphi}(N), \boldsymbol{\varpi}(N))}\Delta N
    +\mathcal{P}_\phi(N,\boldsymbol{\varphi}(N), \boldsymbol{\varpi}(N)) \Delta W_i, } \\
    &\bmte{\varpi_i(N+\Delta N) = \varpi_i(N) + \left(-3\varpi_i(N)-\frac{\partial_{\phi_i} V(\boldsymbol{\varphi}(N))}{H(\boldsymbol{\varphi}(N), \boldsymbol{\varpi}(N))}\right) \Delta N,}
    \label{eq:SDEDisc}
}
where $\Delta N$ is the discretization step width, and $\Delta W_i$ is a number sampled from the normal distribution with zero mean and variance $\Delta N$.
Also, we neglected $\xi_{\pi_i}$ as they are slow-roll suppressed in general.\footnote{The noise $\xi_{\pi_i}$ can be non-negligible beyond the slow-roll cases, such as \ac{USR} inflation represented by Starobinsky's linear-potential model we discuss in Sec.~\ref{sec:Staro}.
For example, Ref.~\cite{Fujita:2025imc} shows that $\pi$'s fluctuation gives an important contribution around the dip of the power spectrum in the ordinary $\delta N$ formalism. However, the paper also shows that it can be neglected around the peak of the power spectrum. Similar results can be seen in the context of the stochastic inflation~\cite{Mizuguchi:2024kbl}. Therefore, we can safely drop the $\xi_{\pi_i}$ term for our purpose.}

Inflation occurs when the inflatons roll in the flat part of the potential, and the potential is dominant in the energy density, ending when the potential is no longer dominant.
According to the $\delta N$ formalism~\cite{Starobinsky:1985ibc,Salopek:1990jq,Sasaki:1995aw,Wands:2000dp,Lyth:2004gb}, the curvature perturbation is related to the fluctuation of the e-fold number that passed during inflation.
Determining a hypersurface $\mathcal{S}_\mathrm{EOI}$ in $\mathbb{R}^{2d}$ called the \ac{EOI} surface, we define a random variable $\mathcal{N}_\mathbf{X}$ as a kind of first passage time: along with a path of $\boldsymbol{\mathcal{X}}$ generated by Eq.~\eqref{eq:SDE} with the initial point $\mathbf{X}$, $\mathcal{N}_\mathbf{X}$ is the first time that $\boldsymbol{\mathcal{X}}$ reaches $\mathcal{S}_\mathrm{EOI}$.
We also define
\begin{align}
\delta \mathcal{N}_{\mathbf{X}}\coloneqq\mathcal{N}_{\mathbf{X}}-\left\langle 
\mathcal{N}_{\mathbf{X}} \right\rangle.
\end{align}
Using this, \ac{AV20}~\cite{Ando:2020fjm} gives the formula for the curvature perturbation power spectrum $\mathcal{P}_\zeta(k)$, which represents the magnitude of $\zeta$ on scale $k$, as follows\footnote{Though \ac{AV20}~\cite{Ando:2020fjm} omitted the dependency on $\boldsymbol{\varpi}$ from their expression~(3.12) for simplicity, our phase-space point $\bfX$ in Eq.~\eqref{eq:PSAndo} includes both $\bm{\varphi}$ and $\bm{\varpi}$ as a more general formula.}
\begin{align}
    \mathcal{P}_\zeta(k) \simeq \int\dd{\mathbf{X}} \pdv{P_\mathrm{bk}}{N_\mathrm{bk}} \qty(N_{\mathrm{bk}(k)},\mathbf{X})\expval{\delta \mathcal{N}_{\mathbf{X}}^2}.
    \label{eq:PSAndo}
\end{align}
Here, $P_\mathrm{bk}(N_\mathrm{bk},\cdot)$ is the \ac{PDF} at backward e-fold $N_\mathrm{bk}$, that is, the \ac{PDF} of $\boldsymbol{\mathcal{X}}$ at $N_\mathrm{bk}$ e-folds before the \ac{EOI}.
$N_{\mathrm{bk}}(k)\coloneqq-\log \left(\frac{k}{k_{\mathrm{end}}}\right)$ is the backward e-fold at which the scale $k$ exits the coarse-graining radius, and $k_{\rm end}\coloneqq k_\sigma(N_{\rm end})$ denotes the coarse-graining scale at the \ac{EOI}. 
Hereafter, we regard $\mathcal{P}_\zeta$ as a function of the backward e-fold $N_\mathrm{bk}$ as usually done in the context of the stochastic-$\delta\calN$ formalism (see, e.g., Refs.~\cite{Fujita:2014tja,Kawasaki:2015ppx,Tada:2023fvd}), letting $\mathcal{P}_\zeta(N_\mathrm{bk})$ be the value of Eq.~\eqref{eq:PSAndo} with $N_{\mathrm{bk}}(k)$ replaced with $N_\mathrm{bk}$.

\subsection{Least squares curve fitting with noisy data \label{sec:LSFit}}

Fitting a dataset of multiple variables to learn the underlying function that describes their relationships is a ubiquitous problem in various fields~ \cite{gyorfi2006distribution}.
In the latter part of this paper, we will consider a kind of such problem.
Concretely, the problem of our interest is formulated as follows: let $X$ be a real-valued random variable and $Y$ be another random variable given as $Y=m(X)+E$, where $m:\mathbb{R}\rightarrow\mathbb{R}$ is an unknown function and $E$ is the noise, that is, a real random variable with zero mean conditioned on $X$. 
Then, given a dataset consisting of $N_{\rm samp}$ independent samples $(X_1,Y_1),\ldots,(X_{N_{\rm samp}},Y_{N_{\rm samp}})$ from the joint probability distribution $\mu_{X,Y}$ of $(X,Y)$, find an approximation function $\tilde{m}$ of $m$.

A common approach to this is least-squares fitting.
We choose a family of functions $f:\mathbb{R}\times\mathbb{R}^{N_{\rm param}}\rightarrow\mathbb{R}$, where the first argument $x$ corresponds to $X$, the second one is a set of $N_{\rm param}$ tunable real parameters denoted by $\boldsymbol{\theta}=(\theta_1,\ldots,\theta_{N_{\rm param}})$, and $f$ is sufficiently smooth with respect to every argument.
Then, we minimize the sum of squared residuals
\begin{align}
    \boldsymbol{\theta}_{\rm min} \coloneqq \argmin_{\boldsymbol{\theta}} \frac{1}{N_{\rm samp}} \sum_{n=1}^{N_{\rm samp}} \left(f(X_n,\boldsymbol{\theta})-Y_n\right)^2,
    \label{eq:thetamin}
\end{align}
and let $\tilde{m}\coloneqq f(\cdot,\boldsymbol{\theta}_{\rm min})$ be an approximation of $m$.
In this approach, even though $Y$ has the noise $E$, we can obtain a good approximation function if we have a large number of data points and a suitable functional form.
Under some assumptions,\footnote{To be precise, although Theorem 11.5 in Ref.~\cite{gyorfi2006distribution} assumes the boundedness of $Y$, that in the problem considered later, which is defined in Eq.~\eqref{eq:OurY}, is not bounded. Hereafter, we simply assume that this does not affect the accuracy of the least squares approximation, assuming that for some sufficiently large $L>0$, the probability that $|Y|>L$ is negligible. Besides, we do not consider the truncation of the approximation function, which is done in Theorem 11.5 in Ref.~\cite{gyorfi2006distribution}, assuming that we obtain a well-fitted function.} the expected error in the form of the averaged squared residual is bounded as follows:
\bae{
    \Mean\left[\int (\tilde{m}(x)-m(x))^2 \mu_X(\dd{x}) \right] \le c\frac{\ln n}{n} + 2 \inf_{\boldsymbol{\theta}} \int (f(x,\boldsymbol{\theta})-m(x))^2 \dd{\mu_X(x)},
}
where $\Mean[{}\cdot{}]$ denotes the expectation with respect to the randomness of samples, $\mu_X$ denotes the distribution of $X$, and $c$ is a constant independent of $n$.
For the details, see Theorem 11.5 in Ref.~\cite{gyorfi2006distribution}.
The first term corresponds to the statistical error, whose decay rate, except for the logarithmic factor, is of order $O(1/n)$ as is common in the Monte Carlo method.
The second term corresponds to the fitting error, and to suppress this term, we need to choose a function family that fits the target function $m$ well. 

The best-fit parameter set $\boldsymbol{\theta}_{\rm min}$ estimated as Eq.~\eqref{eq:thetamin} is a random variable due to the randomness of the dataset $\{(X_n,Y_n)\}_{n=1,\ldots,N_{\rm samp}}$. 
Under the linear approximation of $f$ as a function of $\boldsymbol{\theta}$ around $\boldsymbol{\theta}=\boldsymbol{\theta}_*$, where
\begin{align}
\boldsymbol{\theta}_* \coloneqq \argmin_{\boldsymbol{\theta}} \int\left(f(x,\boldsymbol{\theta})-y\right)^2 \dd{\mu_{X,Y}(x,y)},    
\end{align}
the covariance matrix $C$ of $\boldsymbol{\theta}_{\rm min}=(\theta_{\mathrm{min},1},\ldots,\theta_{\mathrm{min},N_{\rm param}})$ becomes~\cite{hansen2013least}
\begin{align}
    C_{ij}=\Cov[\theta_{\mathrm{min},i},\theta_{\mathrm{min},j}] \simeq \sigma_{\rm res}^2 \left((J^T J)^{-1}\right)_{ij}.
\end{align}
Here, $J$ is the Jacobian matrix of $f$, a $N_{\rm samp} \times N_{\rm param}$ matrix with the $(n,i)$-th entry $J_{ni}=\frac{\partial f}{\partial \theta_i}(X_n,\boldsymbol{\theta}_*)$, and $\sigma_{\rm res}^2$ is the variance of $f(X,\boldsymbol{\theta}_*)-Y$.
In practice, $\boldsymbol{\theta}_*$ is unknown, and so we approximate $J$ as $\tilde{J}$ with the $(n,i)$-th entry $\tilde{J}_{ni}\simeq\frac{\partial f}{\partial \theta_i}(X_n,\boldsymbol{\theta}_{\rm min})$ and $\sigma_{\rm res}^2$ as the sample variance $\tilde{\sigma}_{\rm res}^2$ of $f(X_n,\boldsymbol{\theta}_{\rm min})-Y_n$, assuming that the sample number is sufficiently large and $\boldsymbol{\theta}_{\rm min}$ is close to $\boldsymbol{\theta}_*$.

Given the approximate parameter covariance matrix $\tilde{C}\coloneqq \tilde{\sigma}_{\rm res}^2(\tilde{J}^T \tilde{J})^{-1}$, we can estimate the error in the obtained approximating function $\tilde{m}(x)$ at a given point $x$ as
\begin{align}
    \sqrt{\Var[\tilde{m}(x)]} \simeq{}& \Delta m(x) \nonumber \\
    \coloneqq{}&\sqrt{\left(\nabla_{\boldsymbol{\theta}}f(x,\boldsymbol{\theta}_{\rm min})\right)^T \cdot \tilde{C} \cdot \nabla_{\boldsymbol{\theta}}f(x,\boldsymbol{\theta}_{\rm min})},
    \label{eq:LSFitError}
\end{align}
where $\nabla_{\boldsymbol{\theta}}f(x,\boldsymbol{\theta}_{\rm min})$ is a $N_{\rm param}$-dimensional column vector with $i$-th entry $\frac{\partial f}{\partial \theta_i}(x,\boldsymbol{\theta}_{\rm min})$.
Then, $\Delta m(x)$ is an estimate of the pointwise error of $\tilde{m}$, under the assumption that $\boldsymbol{\theta}_{\rm min}$ is sufficiently close to $\boldsymbol{\theta}_*$ and the fitting error is negligible.

In estimating the power spectrum of the inflationary perturbation, which we consider later, we also aim to obtain the approximation function $\tilde{m}^\prime$ of $m^\prime$, the derivative of $m$.
If $f$ is differentiable with respect to the first argument $x$, we obtain $\tilde{m}^\prime$ just by differentiating $\tilde{m}$, that is, $\tilde{m}^\prime\coloneqq \frac{\partial f}{\partial x}(\cdot,\boldsymbol{\theta}_{\rm min})$.
The error in $\tilde{m}^\prime$ is also estimated similarly to the above as
\begin{align}
    \Delta m^\prime(x)=\sqrt{\left(\nabla_{\boldsymbol{\theta}}\frac{\partial f}{\partial x}(x,\boldsymbol{\theta}_{\rm min})\right)^T \cdot \tilde{C} \cdot \nabla_{\boldsymbol{\theta}}\frac{\partial f}{\partial x}(x,\boldsymbol{\theta}_{\rm min})}.
    \label{eq:LSFitErrorDeriv}
\end{align}


\section{Proposed method \label{sec:method}}

\subsection{Naive method by nested Monte Carlo simulation}

Although $\mathcal{P}_\zeta$ is formally given by Eq.~\eqref{eq:PSAndo}, evaluating it is another issue.
\ac{AV20}~\cite{Ando:2020fjm} gives analytic evaluations in some simple inflation models, but in general models, especially multifield ones, we need to numerically calculate $\mathcal{P}_\zeta$.
Seeing Eq.~\eqref{eq:PSAndo} that involves the probability and the expectation, we naturally consider that we can use the Monte Carlo method to calculate it.
However, naively applying the Monte Carlo method has an issue, which we now see before presenting the method we propose.

\begin{algorithm}
    \caption{The naive method for calculating the power spectrum by the nested Monte Carlo simulation.}
    \label{alg:naive}
    \begin{algorithmic}[1]
    \REQUIRE \ \\
    \begin{itemize}
        \item $n_{\rm path,1}$: the number of sample paths
        \item $n_{\rm path,2}$: the number of branched sample paths
        \item $\mathbf{X}_\mathrm{ini}$: the initial values of $\boldsymbol{\mathcal{X}}$
        \item $N_{\mathrm{bk,1}},\ldots,N_{\mathrm{bk},n_\mathrm{PS}}$: $n_\mathrm{PS}$ values of $N_{\mathrm{bk}}$ at which $\mathcal{P}_\zeta(N_\mathrm{bk})$ is calculated
        \item $\Delta N_\mathrm{bk}$: the width for the finite difference approximation of $N_\mathrm{bk}$ derivative
    \end{itemize}

    \ENSURE  
    $\tilde{\calP}_{\zeta,i}$: approximate $\calP_\zeta(N_{\bk,i})$ for $i=1,\dots,n_\PS$.

    \FOR{$n=1,...,n_{\rm path,1}$}

    \STATE Generate a path $\omega_n$ from $\mathbf{X}_\mathrm{ini}$ to the \ac{EOI} based on Eq.~\eqref{eq:SDEDisc}.

    \FOR{$m=1,...,n_{\rm PS}$}
    
    \STATE Let the values of $\boldsymbol{\mathcal{X}}$ at the backward e-folds $N_{\mathrm{bk},m}\pm\Delta N_\mathrm{bk}$ on $\omega_n$ be $\mathbf{X}_{\mathrm{bk},n,m,\pm}$, respectively.

    \FOR{$l=1,...,n_{\rm path,2}$}

    \STATE Generate a path from $\mathbf{X}_{\mathrm{bk},n,m,+}$ (resp. $\mathbf{X}_{\mathrm{bk},n,m,-}$) to the \ac{EOI} and let the elapsed e-fold $N_{n,m,l,+}$ (resp. $N_{n,m,l,-}$).

    \ENDFOR

    \STATE Compute
    \beae{
        V_{n,m,\pm}\coloneqq \bmte{
        \frac{1}{n_{\rm path,2}} \sum_{l=1}^{n_{\rm path,2}} N_{n,m,l,\pm}^2 
        - \left(\frac{1}{n_{\rm path,2}} \sum_{l=1}^{n_{\rm path,2}} N_{n,m,l,\pm}\right)^2.
        }
    }

    \ENDFOR
    \ENDFOR

    \FOR{$m=1,...,n_{\rm PS}$}

    \STATE Compute $F^\pm_m \coloneqq \frac{1}{n_{\rm path,1}} \sum_{n=1}^{n_{\rm path,1}} V_{n,m,\pm}$.

    \STATE Output $\tilde{\mathcal{P}}_{\zeta,m}\coloneqq \frac{1}{2\Delta N_\mathrm{bk}}(F^+_m - F^-_m)$

    \ENDFOR
    
    \end{algorithmic}
\end{algorithm}

We see that
\begin{align}
    \mathcal{P}_\zeta(N_\mathrm{bk}) \simeq \dv{F_{\langle\delta \mathcal{N}^2\rangle}}{N_\mathrm{bk}}{}(N_\mathrm{bk}),
\end{align}
where we define
\begin{align}
    F_{\langle\delta \mathcal{N}^2\rangle}(N_\mathrm{bk}) \coloneqq \int \dd{\mathbf{X}} P_\mathrm{bk}\left(N_\mathrm{bk},\mathbf{X}\right) \left\langle \delta \mathcal{N}_{\mathbf{X}}^2 \right\rangle.
    \label{eq:FdelN2Sq}
\end{align}
Approximating the $N_\mathrm{bk}$ derivative by the finite difference yields
\bae{
    \mathcal{P}_\zeta(N_\mathrm{bk}) \simeq \frac{1}{2\Delta N_\mathrm{bk}}\left(F_{\langle\delta \mathcal{N}^2\rangle}(N_\mathrm{bk}+\Delta N_\mathrm{bk})- F_{\langle\delta \mathcal{N}^2\rangle}(N_\mathrm{bk}-\Delta N_\mathrm{bk})\right),
    \label{eq:PzetaFD}
}
where the small displacement $\Delta N_\mathrm{bk}$ is set appropriately.
The problem thus boils down to calculating $F_{\langle\delta \mathcal{N}^2\rangle}(N_\mathrm{bk})$ for given $N_\mathrm{bk}$.
We note that this is a nested expectation calculation: $F_{\langle\delta \mathcal{N}^2\rangle}(N_\mathrm{bk})$ is the expectation of $\left\langle \delta \mathcal{N}_{\mathbf{X}}^2 \right\rangle$ with $\mathbf{X}$ obeying the \ac{PDF} $P_\mathrm{bk}\left(N_\mathrm{bk},\cdot\right)$, and $\left\langle \delta \mathcal{N}_{\mathbf{X}}^2 \right\rangle$ is also an expectation.
Then, we conceive the Monte Carlo-based algorithm~\ref{alg:naive} to calculate $\mathcal{P}_\zeta(N_\mathrm{bk})$.

A brief explanation of this algorithm is as follows. 
Note that $\mathbf{X}_{\mathrm{bk},n,m,\pm}$ in line~4 is a sample from the \ac{PDF} $P_\mathrm{bk}\left(N_\mathrm{bk}\pm\Delta N_\mathrm{bk},\cdot\right)$.
With this being $\mathbf{X}$, we generate samples of $\mathcal{N}_{\mathbf{X}}$ in line 6 and calculate $\left\langle \delta \mathcal{N}_{\mathbf{X}}^2 \right\rangle$ as a sample variance in line~8.
Then, in line~12, we compute the average of these and let it be an approximation of $F_{\langle\delta \mathcal{N}^2\rangle}(N_\mathrm{bk}\pm\Delta N_\mathrm{bk})$.
This approach was, in fact, taken in the previous paper~\cite{Fujita:2014tja}, where the author used a formula for $\mathcal{P}_{\zeta}$ slightly different from Eq.~\eqref{eq:PSAndo}.

Although Algorithm~\ref{alg:naive} is straightforward, it has some shortcomings.
First, it is a nested Monte Carlo simulation and thus takes a long computational time.
In the algorithm, updating $\boldsymbol{\mathcal{X}}$ by Eq.~\eqref{eq:SDEDisc} is the computation that is iterated the most and is the most time-consuming.
The total number of updates is evaluated as
\begin{equation}
    O\left(n_{\mathrm{path},1}n_{\mathrm{path},2}n_{\rm PS}n_t\right),
    \label{eq:compNaive}
\end{equation}
where $n_t \coloneqq N_{\rm tot} / \Delta N_{\rm bk}$ is the typical number of time steps in a path and $N_{\rm tot}$ is the typical e-fold that elapses from the initial point to the \ac{EOI}.
Eq.~\eqref{eq:compNaive} implies that the computational time becomes very long since it is proportional to the product of $n_{\mathrm{path},1}$ and $n_{\mathrm{path},2}$, which, to suppress the statistical error in Monte Carlo simulation, we need to set large.
According to the well-known fact about the Monte Carlo method that for the statistical error at most $\epsilon$, the required sample number is of order $O(1/\epsilon^2)$, we need to take $n_{\mathrm{path},1},n_{\mathrm{path},2}=O(1/\epsilon^2)$, which makes Eq.~\eqref{eq:compNaive} $O(1/\epsilon^4)$.
In fact, in Ref.~\cite{Fujita:2014tja}, both of them were set to $10^4$, whose product becomes $10^8$.

Another issue is that, as we see from Eq.~\eqref{eq:compNaive} proportional to $n_{\rm PS}$, the computational time becomes long if we want $\mathcal{P}_\zeta(N_{\rm bk})$ for many values of $N_{\rm bk}$, which is required for precise discussion on cosmological implications of the calculated power spectrum.
Moreover, along with the statistical error, $\mathcal{P}_\zeta$ calculated by Eq.~\eqref{eq:PzetaFD} contains the error by the finite difference approximation.

As we will see in the following, the method we propose solves these shortcomings of Algorithm \ref{alg:naive}.

\subsection{Method by unnested Monte Carlo simulation}

Now, let us see that we can avoid the nested Monte Carlo simulation by rewriting the formula~\eqref{eq:FdelN2Sq} for $F_{\langle\delta \mathcal{N}\rangle^2}$.
By a simple calculation, we see that
\begin{align}
    \left\langle \delta \mathcal{N}_{\mathbf{X}}^2 \right\rangle= \left\langle\frac{1}{2}\left(\mathcal{N}_{\mathbf{X}}^{(1)}-\mathcal{N}_{\mathbf{X}}^{(2)}\right)^2\right\rangle,
    \label{eq:delNSqEstimator}
\end{align}
where $\mathcal{N}_{\mathbf{X}}^{(1)}$ and $\mathcal{N}_{\mathbf{X}}^{(2)}$ are two i.i.d. random variables with the same distribution as $\mathcal{N}_{\mathbf{X}}$.
Thus, we can write Eq.~\eqref{eq:FdelN2Sq} as
\begin{align}
    & F_{\langle\delta \mathcal{N}^2\rangle}(N_\mathrm{bk}) 
    \nonumber \\
    & \bmbe{=\int \dd{\mathbf{X}} \int_0^\infty \dd{N_{\rm bk,1}} \int_0^\infty \dd{N_{\rm bk,2}} P_\mathrm{bk}\left(N_\mathrm{bk},\mathbf{X}\right)
    P_{\mathcal{N}_\mathbf{X}}(N_{\rm bk,1}) P_{\mathcal{N}_\mathbf{X}}(N_{\rm bk,2})\frac{1}{2}(N_{\rm bk,1}-N_{\rm bk,2})^2,}
    \label{eq:FdelN2Sq2}
\end{align}
where $P_{\mathcal{N}_\mathbf{X}}$ is the \ac{PDF} of $\mathcal{N}_{\mathbf{X}}$.

Having this formula, we conceive another Monte Carlo-based approach to calculate $F_{\langle\delta \mathcal{N}^2\rangle}(N_\mathrm{bk})$: we sample many tuples $(\mathbf{X},N_{\rm bk,1},N_{\rm bk,2})$ from the joint \ac{PDF} $P_{\rm joint}(\mathbf{X},N_{\rm bk,1},N_{\rm bk,2})\coloneqq P_\mathrm{bk}\left(N_\mathrm{bk},\mathbf{X}\right)P_{\mathcal{N}_\mathbf{X}}(N_{\rm bk,1}) P_{\mathcal{N}_\mathbf{X}}(N_{\rm bk,2})$ and take the sample average of $\frac{1}{2}(N_{\rm bk,1}-N_{\rm bk,2})^2$ as an approximation of $F_{\langle\delta \mathcal{N}^2\rangle}(N_\mathrm{bk})$.
The concrete procedure is given as Algorithm \ref{alg:unnest}.
%
\begin{algorithm}
    \caption{The Monte Carlo-based method for calculating the power spectrum without the nested simulation.}
    \label{alg:unnest}
    \begin{algorithmic}[1]
    \REQUIRE \ \\
    \begin{itemize}
        \item $n_{\rm path}$: the number of sample paths
        \item $\mathbf{X}_\mathrm{ini}$: the initial values of $\boldsymbol{\mathcal{X}}$
        \item $N_{\mathrm{bk,1}},\ldots,N_{\mathrm{bk},n_\mathrm{PS}}$: $n_\mathrm{PS}$ values of $N_{\mathrm{bk}}$ at which $\mathcal{P}_\zeta(N_\mathrm{bk})$ is calculated
        \item $\Delta N_\mathrm{bk}$: the width for the finite difference approximation of $N_\mathrm{bk}$ derivative
    \end{itemize}

    \ENSURE $\tilde{\calP}_{\zeta,i}$: approximate $\calP_\zeta(N_{\bk,i})$ for $i=1,\dots,n_\PS$.

    \FOR{$n=1,...,n_{\rm path}$}

    \STATE Generate a path $\omega_n$ from $\mathbf{X}_\mathrm{ini}$ to the \ac{EOI} based on Eq.~\eqref{eq:SDEDisc}.

    \FOR{$m=1,...,n_{\rm PS}$}
    
    \STATE Let the values of $\boldsymbol{\mathcal{X}}$ at the backward e-folds $N_{\mathrm{bk},m}\pm\Delta N_\mathrm{bk}$ on $\omega_n$ be $\mathbf{X}_{\mathrm{bk},n,m,\pm}$, respectively.

    \STATE Generate two paths from $\mathbf{X}_{\mathrm{bk},n,m,+}$ (resp. $\mathbf{X}_{\mathrm{bk},n,m,-}$) to the \ac{EOI} and let the elapsed e-folds be $N_{n,m,1,+}$ and $N_{n,m,2,+}$ (resp. $N_{n,m,1,-}$ and $N_{n,m,2,-}$).

    \ENDFOR
    \ENDFOR

    \FOR{$m=1,...,n_{\rm PS}$}

    \STATE Compute 
    \bae{\tilde{F}^\pm_m \coloneqq \frac{1}{n_{\rm path}} \sum_{n=1}^{n_{\rm path}} \frac{1}{2}\left(N_{n,m,1,\pm}-N_{n,m,2,\pm}\right)^2.}

    \STATE Output $\tilde{\mathcal{P}}_{\zeta,m}\coloneqq \frac{1}{2\Delta N_\mathrm{bk}}(F^+_m - F^-_m)$

    \ENDFOR
    \end{algorithmic}
\end{algorithm}
This method avoids the nested Monte Carlo simulation, and its computational time is
\begin{align}
    O(n_{\rm path}n_{\rm PS}n_t).
\end{align}
However, it still gives the power spectrum only for preselected discrete values of $N_{\rm bk}$ and takes time proportional to $n_{\rm PS}$, the number of values of $N_{\rm bk}$.
Additionally, the finite difference error persists.

\subsection{Method by Monte Carlo simulation and least squares curve fitting}

We now present our main proposal, which involves calculating the power spectrum using a Monte Carlo simulation and least squares curve fitting.

\begin{algorithm}
    \caption{The proposed method for finding an approximation function of the power spectrum by Monte Carlo simulation and least squares fitting (MCLSFit).}
    \label{alg:main}
    \begin{algorithmic}[1]

    \REQUIRE \ \\
    \begin{itemize}
        \item $n_{\rm path}$: the number of sample paths
        \item $\mathbf{X}_\mathrm{ini}$: the initial values of $\boldsymbol{\mathcal{X}}$
        \item $[N_{\mathrm{bk,l}},N_{\mathrm{bk,u}}]$: the range of the backward e-fold on which the power spectrum is approximated
        \item $f(N_{\mathrm{bk}},\boldsymbol{\theta})$: the real-valued function that has the parameters $\boldsymbol{\theta}$ and is differentiable with $N_{\mathrm{bk}}$
    \end{itemize}

    \ENSURE Functions $\tilde{F}_{\left\langle \delta N^2 \right\rangle}$ and $\tilde{\mathcal{P}}_{\zeta}$ that approximate $F_{\left\langle \delta N^2 \right\rangle}$ and $\mathcal{P}_{\zeta}$, respectively.

    \FOR{$n=1,...,n_{\rm path}$}

    \STATE Sample a value $N_{\mathrm{bk},n}$ from $U(N_{\mathrm{bk,l}},N_{\mathrm{bk,u}})$, the uniform distribution with support $[N_{\mathrm{bk,l}},N_{\mathrm{bk,u}}]$.

    \STATE Generate a path $\omega_n$ from $\mathbf{X}_\mathrm{ini}$ to the \ac{EOI} based on Eq.~\eqref{eq:SDEDisc}.
    
    \STATE Let the value of $\boldsymbol{\mathcal{X}}$ at the backward e-fold $N_{\mathrm{bk},n}$ on $\omega_n$ be $\mathbf{X}_{\mathrm{bk},n}$.

    \STATE Generate two paths from $\mathbf{X}_{\mathrm{bk},n}$ to the \ac{EOI} and let the elapsed e-folds be $N_{n,1}$ and $N_{n,2}$.

    \ENDFOR

    \STATE
    Find
    \begin{equation}
        \boldsymbol{\theta}_{\mathrm{min}}\coloneqq\argmin_{\boldsymbol{\theta}} \sum_{n=1}^{n_\mathrm{path}} \left[f(N_{\mathrm{bk},n},\boldsymbol{\theta})-\frac{1}{2}\left(N_{n,1}-N_{n,2}\right)^2\right]^2.
    \end{equation}

    \STATE Output $f(\cdot,\boldsymbol{\theta}_{\mathrm{min}})$ and $\partial_{N_{\mathrm{bk}}}f(\cdot,\boldsymbol{\theta}_{\mathrm{min}})$ as $\tilde{F}_{\left\langle \delta N^2 \right\rangle}$ and $\tilde{\mathcal{P}}_{\zeta}$, respectively.

    \end{algorithmic}
\end{algorithm}

We aim to find a function $\tilde{F}_{\langle\delta \mathcal{N}^2\rangle}(N_\mathrm{bk})$ that well fits $F_{\langle\delta \mathcal{N}^2\rangle}(N_\mathrm{bk})$ over some range $[N_{\rm bk,l},N_{\rm bk,u}]$ of $N_{\rm bk}$ that we are interested in.
To do so, we consider this task as a kind of problem described in Sec.~\ref{sec:LSFit}.
We view a random variable $\mathcal{N}_{\rm bk}$ that uniformly distributes in $[N_{\rm bk,l},N_{\rm bk,u}]$ as $X$ and
\begin{align}
    Y_{\langle\delta \mathcal{N}^2\rangle}\coloneqq\frac{1}{2}\left(\mathcal{N}_{\boldsymbol{\mathcal{X}}_{\rm bk}}^{(1)}-\mathcal{N}_{\boldsymbol{\mathcal{X}}_{\rm bk}}^{(2)}\right)^2.
    \label{eq:OurY}
\end{align}
as $Y$.
Here, $\boldsymbol{\mathcal{X}}_{\rm bk}$ is a $\mathbb{R}^{2d}$-valued random variable such that the joint distribution of $(\mathcal{N}_{\rm bk},\boldsymbol{\mathcal{X}}_{\rm bk})$ has the \ac{PDF}
\begin{align}
P_{\mathcal{N}_{\rm bk},\boldsymbol{\mathcal{X}}_{\rm bk}}(N_{\rm bk},\mathbf{X}_{\rm bk})=P_{\mathcal{N}_{\rm bk}}(N_{\rm bk})P_\mathrm{bk}\left(N_\mathrm{bk},\mathbf{X}_{\rm bk}\right),
\end{align}
where
\begin{align}
    P_{\mathcal{N}_{\rm bk}}(N_{\rm bk})\coloneqq
    \begin{dcases}
        \frac{1}{N_{\rm bk,u}-N_{\rm bk,l}} & ; \ N_{\rm bk,l} \le N_{\rm bk} \le N_{\rm bk,u} \\
        0 & ; \ \mathrm{otherwise}
    \end{dcases}
\end{align}
is the \ac{PDF} of $\mathcal{N}_{\rm bk}$.
$\mathcal{N}_{\boldsymbol{\mathcal{X}}_{\rm bk}}^{(1)}$ and $\mathcal{N}_{\boldsymbol{\mathcal{X}}_{\rm bk}}^{(2)}$ are i.i.d. random variables, each of which, conditioned that $\boldsymbol{\mathcal{X}}_{\rm bk}=\mathbf{X}_{\rm bk}$, has the same \ac{PDF} as $\mathcal{N}_{\mathbf{X}_{\rm bk}}$.
Now, we decompose $Y$ as $Y=m(\mathcal{N}_\mathrm{bk})+E$ with
\bae{
    m(\mathcal{N}_\mathrm{bk})=F_{\langle\delta \mathcal{N}^2\rangle}(\mathcal{N}_\mathrm{bk}), \quad 
    E=\frac{1}{2}\left(\mathcal{N}_{\boldsymbol{\mathcal{X}}_{\rm bk}}^{(1)}-\mathcal{N}_{\boldsymbol{\mathcal{X}}_{\rm bk}}^{(2)}\right)^2 - F_{\langle\delta \mathcal{N}^2\rangle}(\mathcal{N}_\mathrm{bk}).
}
By definition,
\begin{equation}
    \Mean\left[E \relmiddle{|}\mathcal{N}_\mathrm{bk}=N_{\rm bk}\right] = 0,
\end{equation}
holds.
Thus, the current task matches the problem setting in Sec.~\ref{sec:LSFit}, and we can find an approximating function $\tilde{F}_{\langle\delta \mathcal{N}^2\rangle}(N_\mathrm{bk})$ of $F_{\langle\delta \mathcal{N}^2\rangle}(N_\mathrm{bk})$ by the method described there: we generate sample values $\left\{\left(N_{{\rm bk},n},Y_{\langle\delta \mathcal{N}^2\rangle,n}\right)\right\}_{n=1,\ldots,n_{\rm path}}$ of $\left(\mathcal{N}_{\rm bk},Y_{\langle\delta \mathcal{N}^2\rangle}\right)$ and perform the least squares fit to the sample set.
The concrete procedure, which we hereafter call MCLSFit, is described in Algorithm~\ref{alg:main}.

The resultant function $\tilde{F}_{\langle\delta \mathcal{N}^2\rangle}$ satisfies that
\bme{
    \Mean\left[\int_{N_{\rm bk,l}}^{N_{\rm bk,u}} \left(\tilde{F}_{\langle\delta \mathcal{N}^2\rangle}(N_{\rm bk})-F_{\langle\delta \mathcal{N}^2\rangle}(N_{\rm bk})\right)^2  \frac{{\rm d}N_{\rm bk}}{N_{\rm bk,u}-N_{\rm bk,l}} \right] \\
    \le c\frac{\ln n_{\rm path}}{n_{\rm path}} + 2 \inf_{\boldsymbol{\theta}} \int_{N_{\rm bk,l}}^{N_{\rm bk,u}} \left(f(N_{\rm bk},\boldsymbol{\theta})-F_{\langle\delta \mathcal{N}^2\rangle}(N_{\rm bk})\right)^2 \frac{{\rm d}N_{\rm bk}}{N_{\rm bk,u}-N_{\rm bk,l}}.
}
We now avoid not only the nested Monte Carlo simulation but also the $n_{\rm PS}$-times iterations: the computational time is
\begin{align}
    O(n_{\rm path}n_t).
\end{align}
We have a function $\tilde{F}_{\langle\delta \mathcal{N}^2\rangle}$ that approximately gives $F_{\langle\delta \mathcal{N}^2\rangle}$ for any $N_{\rm bk}\in[N_{\rm bk,l},N_{\rm bk,u}]$, not the values at a finite number of preselected $N_{\rm bk}$.
Besides, if $f(N_{\mathrm{bk}},\boldsymbol{\theta})$ is differentiable with respect to $N_{\mathrm{bk}}$, we also have an approximation function $\tilde{P}_\zeta=\dv*{\tilde{F}_{\langle\delta \mathcal{N}^2\rangle}}{N_{\rm bk}}$ of the power spectrum $P_\zeta$, which does not rely on the finite difference approximation.
Moreover, if $f(N_{\mathrm{bk}},\boldsymbol{\theta})$ is differentiable with respect to $\boldsymbol{\theta}$, we can estimate the error in $\tilde{F}_{\langle\delta \mathcal{N}^2\rangle}$ and $\tilde{P}_\zeta$ as described in Sec.~\ref{sec:LSFit}.

An issue is choosing a suitable function $f(N_{\mathrm{bk}},\boldsymbol{\theta})$ that can be fitted to $F_{\left\langle \delta N^2 \right\rangle}$ well.
As is common in any machine learning problem, we should have some prior knowledge about the fitted function $F_{\left\langle \delta N^2 \right\rangle}$ in certain ways, such as qualitative discussions and approximate analyses, and reflect it onto $f$.
In fact, in some of the test cases considered in Sec.~\ref{sec:result}, we use qualitative knowledge to choose $f$.
If we do not have enough knowledge, the following way using the samples $\{(N_{{\rm bk},n},N_{n,1},N_{n,2})\}_n$ would be helpful: we bin the samples based on $N_{\rm bk}$ and take an average of $\frac{1}{2}(N_1-N_2)^2$ in each bin.
This helps us roughly grasp the shape of $\mathcal{P}_\zeta$ and get guidance on choosing the fitting function.
This can also be used to check the fitting quality afterward, even if we have plausible fitting functions.
The concrete procedure of this method, which we hereafter call MCBinAve, is as shown in Algorithm~\ref{alg:binAve}.

\begin{algorithm}
    \caption{The proposed method for finding an approximation of the power spectrum by binning and averaging the Monte Carlo samples (MCBinAve).}
    \label{alg:binAve}
    \begin{algorithmic}[1]

    \REQUIRE \ \\
    \begin{itemize}
        \item $n_{\rm path}$: the number of sample paths
        \item $\mathbf{X}_\mathrm{ini}$: the initial values of $\boldsymbol{\mathcal{X}}$
        \item $[N_{\mathrm{bk,l}},N_{\mathrm{bk,u}}]$: the range of the backward e-fold on which the power spectrum is approximated
        \item $n_{\rm bin}$: the number of bins
    \end{itemize}

    \STATE Run steps~1--6 in Algorithm~\ref{alg:main}.

    \STATE Set $N_{\mathrm{bin},m}\coloneqq N_{\mathrm{bk,l}}+m\Delta N_{\rm bin}$ with $\Delta N_{\rm bin}\coloneqq (N_{\mathrm{bk,u}}-N_{\mathrm{bk,l}})/n_{\rm bin}$ for $m=0,\ldots,n_{\rm bin}$.

    \FOR{$m=1,...,n_{\rm bin}$}

    \STATE
    Compute
    \bae{
        \tilde{F}_{m}\coloneqq\frac{1}{M_m}\sum_{n=1}^{n_\mathrm{path}}\frac{1}{2}\left(N_{n,1}-N_{n,2}\right)^2\textbf{1}_{N_{\mathrm{bin},m-1} \le N_{\mathrm{bk},n} < N_{\mathrm{bin},m}},
    }
    where the indicator function $\textbf{1}_C$ takes 1 if the condition $C$ holds and 0 otherwise, and $M_m\coloneqq \sum_{n=1}^{n_\mathrm{path}} \textbf{1}_{N_{\mathrm{bin},m-1} \le N_{\mathrm{bk},n} < N_{\mathrm{bin},m}}$.

    \ENDFOR
    
    \FOR{$m=1,...,n_{\rm bin}-1$}

    \STATE Output $\tilde{\mathcal{P}}_{\zeta,m}\coloneqq \frac{\tilde{F}_{m+1}-\tilde{F}_{m}}{\Delta N_{\rm bin}}$ as an approximation of $\mathcal{P}_{\zeta}(N_{\mathrm{bin},m})$.

    \ENDFOR

    \end{algorithmic}
\end{algorithm}

We can regard $\tilde{F}_m$ in Algorithm~\ref{alg:binAve} as a Monte Carlo estimation of
\begin{align}
\overline{F}(N_{\rm bk})\coloneqq\frac{1}{\Delta N_{\rm bin}}\int_{N_{\rm bk}-\frac{1}{2}\Delta N_{\rm bin}}^{N_{\rm bk}+\frac{1}{2}\Delta N_{\rm bin}} F_{\langle\delta \mathcal{N}^2\rangle}(N_{\rm bk})  {\rm d}N_{\rm bk},
\end{align}
the average of $F_{\langle\delta \mathcal{N}^2\rangle}$ over the bin, and $\tilde{\mathcal{P}}_{\zeta,m}$ as that of
\begin{align}
    \overline{\mathcal{P}}_\zeta(N_{\rm bk})\coloneqq \frac{\overline{F}\left(N_{\rm bk}+\frac{1}{2}\Delta N_{\rm bin}\right)-\overline{F}\left(N_{\rm bk}-\frac{1}{2}\Delta N_{\rm bin}\right)}{\Delta N_{\rm bin}},
\end{align}
the finite-difference approximation of the derivative of $\overline{F}$.
The standard errors of these estimations are calculated as
\bae{
    \Delta \tilde{F}_m \coloneqq \frac{1}{\sqrt{M_m}}\Biggl(\frac{1}{M_m}\sum_{n=1}^{n_\mathrm{path}}\frac{1}{4}\left(N_{n,1}-N_{n,2}\right)^4 \textbf{1}_{N_{\mathrm{bin},m-1} \le N_{\mathrm{bk},n} < N_{\mathrm{bin},m}} -\tilde{F}_m^2\Biggr)^{1/2},
    \label{eq:errBindelNSq}
}
and
\begin{align}
    \Delta \tilde{\mathcal{P}}_{\zeta,m} \coloneqq \frac{\sqrt{\left(\Delta \tilde{F}_m\right)^2+\left(\Delta \tilde{F}_{m+1}\right)^2}}{\Delta N_{\rm bin}}.
    \label{eq:errBinPS}
\end{align}

\subsection{Relation of the adopted power spectrum formula to others}\label{sec: Animali}

Before moving to the concrete examples, we mention the difference in the variants of the power spectrum formula~\eqref{eq:PSAndo} proposed so far in the literature. 
While \ac{AV20} relates the scale of interest $k$ with the phase-space point $\bfX$ by the backward e-folds $N_\bk(k)\coloneqq-\ln(k/k_\uend)$ from \ac{EOI},
the original stochastic-$\delta\calN$ approach by \ac{FKTT}~\cite{Fujita:2013cna,Fujita:2014tja} chose the point $\bfX$ from which the average e-folds to \ac{EOI}, $\expval{\calN(\bfX)}$, equal to $-\ln(k/k_\uend)$.
In the latest formulation by \ac{AV24} where the consistency between the two-point function and the spatial coarse-graining is carefully investigated, it is found that the \ac{FKTT} formula was incidentally not a bad approximation in the large-volume limit $k\ll k_\uend$, though the average should be volume-weighted: the variance $\expval{\delta\calN^2_\bfX}$ should be replaced by 
\bae{
    \expval{\delta\calN^2_\bfX}_V\coloneqq\frac{\expval{e^{3\calN_\bfX}\calN_\bfX^2}}{\expval{e^{3\calN_\bfX}}}-\pqty{\frac{\expval{e^{3\calN_\bfX}\calN_\bfX}}{\expval{e^{3\calN_\bfX}}}}^2,
}
and the scale is related to the point so that
\bae{
    \pqty{\frac{k_\uend}{2k}}^3=\expval{e^{3\calN_\bfX}},
}
where the factor two reflects the fact that $k$ should correspond not to the diameter but to the radius of the patch of interest.
All these formulations of the power spectrum are summarized in the following expression:
\bae{
    \calP_\zeta(k)\simeq\int\dd{\bfX}\pdv{P_k(\bfX)}{(-\ln k)}\doublevev{\delta\calN^2_\bfX}.
}
Here, $P_k(\bfX)$ is the \ac{PDF} of $\bfX$ on a certain hypersurface corresponding to the scale $k$, as summarized by
\bae{
    P_k(\bfX)=\bce{
        P\pqty{\bfX\mid\expval{\calN_\bfX}=-\ln(k/k_\uend)} & \text{for \ac{FKTT}}, \\
        P_\bk(N_\bk(k),\bfX) & \text{for \ac{AV20}}, \\
        P\pqty{\bfX\mid\expval{e^{3\calN_\bfX}}=(k_\uend/(2k))^3} & \text{for \ac{AV24}}.
    }
}
$\doublevev{\delta\calN_\bfX^2}$ represents the standard variance $\expval{\delta\calN_\bfX^2}$ for \ac{FKTT} and \ac{AV20}, and the volume-weighted one $\expval{\delta\calN_\bfX^2}_V$ for \ac{AV24}.
Their differences are higher-order in $\delta\calN$ and hence the curvature perturbation, so they are expected not to be significant if $\calP_\zeta(k)\ll1$ as in a realistic model and most examples we will see below.
Only the flat quantum well model in our examples can achieve $\calP_\zeta>1$ (see Fig.~6 of \ac{AV20}~\cite{Ando:2020fjm}) and exhibits a characteristic ``leakage" feature from small-scale perturbations to large-scale ones.
This feature is indeed predicted only in the \ac{AV20} formula because the surface condition $\expval{\calN_\bfX}=-\ln(k/k_\uend)$ in \ac{FKTT} or $\expval{e^{3\calN_\bfX}}=\pqty{k_\uend/(2k)}^3$ in \ac{AV24} uniquely determines the corresponding inflaton value $\varphi_*$.
We will come back to this point in Sec.~\ref{sec:well}, but we will merely focus on reproducing the \ac{AV20} result in the least squares fitting method without discussing the validity of this feature.

Let us also mention other approaches to computing the power spectrum in the stochastic formalism. One is the adjoint Fokker--Planck approach~\cite{Tada:2023fvd} as a kind of derivative of \ac{FKTT}. There, the average $\expval{\calN_{\bm{\calX}}}$ and the variance $\expval{\delta\calN_{\bm{\calX}}^2}$ are calculated as functions of the phase-space point $\bm{\calX}$ via the adjoint Fokker--Planck \ac{PDE} proposed in Ref.~\cite{Vennin:2015hra} rather than the Monte Carlo way.
While the Monte Carlo approach cannot calculate the variance smaller than the discretization step square, $(\Delta N)^2$, of the Langevin equation (Eq.~\eqref{eq:SDEDisc}), the \ac{PDE} approach is free from such a limitation.
On the other hand, numerical solvers for \acp{PDE} in general suffer from the \emph{curse of dimensionality}, i.e., the exponential complexity in the number of fields $d$ (or $2d$ for phase space).
They are hence complementary to each other.

The lattice implementation of the stochastic formalism, dubbed \emph{\ac{STOLAS}} in Ref.~\cite{Mizuguchi:2024kbl}, is a more direct approach compared to the derivatives of \ac{FKTT}.
As it literally simulates the spatial distribution of the curvature perturbation, the power spectrum (and beyond, such as the bispectrum and trispectrum, in principle) can be obtained by simply Fourier-transforming the real-space distribution.
The calculable e-fold range $N_{\bk,\mathrm{u}}-N_{\bk,\mathrm{l}}$ (within one simulation) is limited by the lattice resolution.
In the latest implementation~\cite{STOLAS}, the single $256^3$-lattice simulation corresponding to $N_{\bk,\mathrm{u}}-N_{\bk,\mathrm{l}}\simeq\ln256\simeq5.5$ takes $10$ minutes in addition to $6$ minutes for noise map generation (by MacBook Pro with Apple M1 Max CPU ($10$ cores), $64$ GB RAM, and no GPU use)).

Another numerical simulation of stochastic inflation is \ac{FOREST} by Ref.~\cite{Animali:2025pyf}.
It computes the spatial correlation of the curvature perturbation as ``stochastic trees", which simulate the branching of Hubble patches.
Though it is not explicitly given in the reference, it is expected to have the ability to calculate the power spectrum.

\section{Numerical results \label{sec:result}}

We now conduct numerical demonstrations to run the proposed algorithms in four test cases.
The first three are single-field models, for each of which the curvature perturbation power spectrum can be obtained at least approximately using existing methods.
Although these models may not require our Monte Carlo-based methods, we can still compare our results with those of existing methods to assess the effectiveness of our approach.
The last one is a multi-field model, for which existing approaches are more challenging than single-field cases, and it is the very situation where our method becomes helpful.

We use \texttt{DifferentialEquations.jl}~\cite{rackauckas2017differentialequations} to generate paths according to Eq.~\eqref{eq:SDEDisc}, adopting the \ac{EM} scheme with time step size $\Delta N=0.01$ and setting the number of paths to $n_{\rm path}=10^5$, unless otherwise specified.
Our code is available at \url{https://github.com/Koichi-Miyamoto/StochasticInflationMCLS}.
All the calculations below were run on Fujitsu LIFEBOOK WP1/J3 with Intel Core Ultra 7 155H CPU (16 cores, 3.0 GHz), 32 GB RAM, and no GPU use.

\subsection{Chaotic inflation}

\begin{figure*} 
    \centering
    \begin{tabular}{c}
        \begin{minipage}{0.48\hsize}
            \centering
            \includegraphics[width=0.95\hsize]{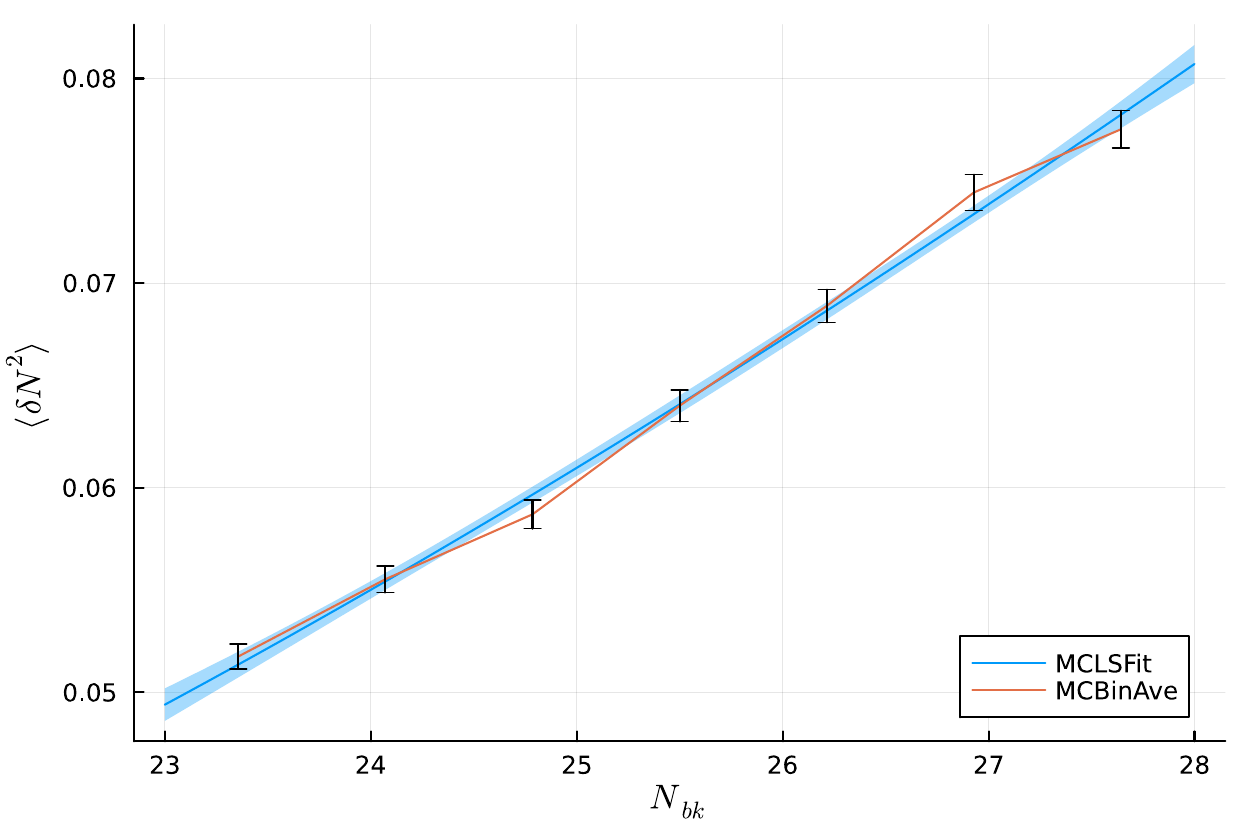}
        \end{minipage}
        \begin{minipage}{0.48\hsize}
            \centering
            \includegraphics[width=0.95\hsize]{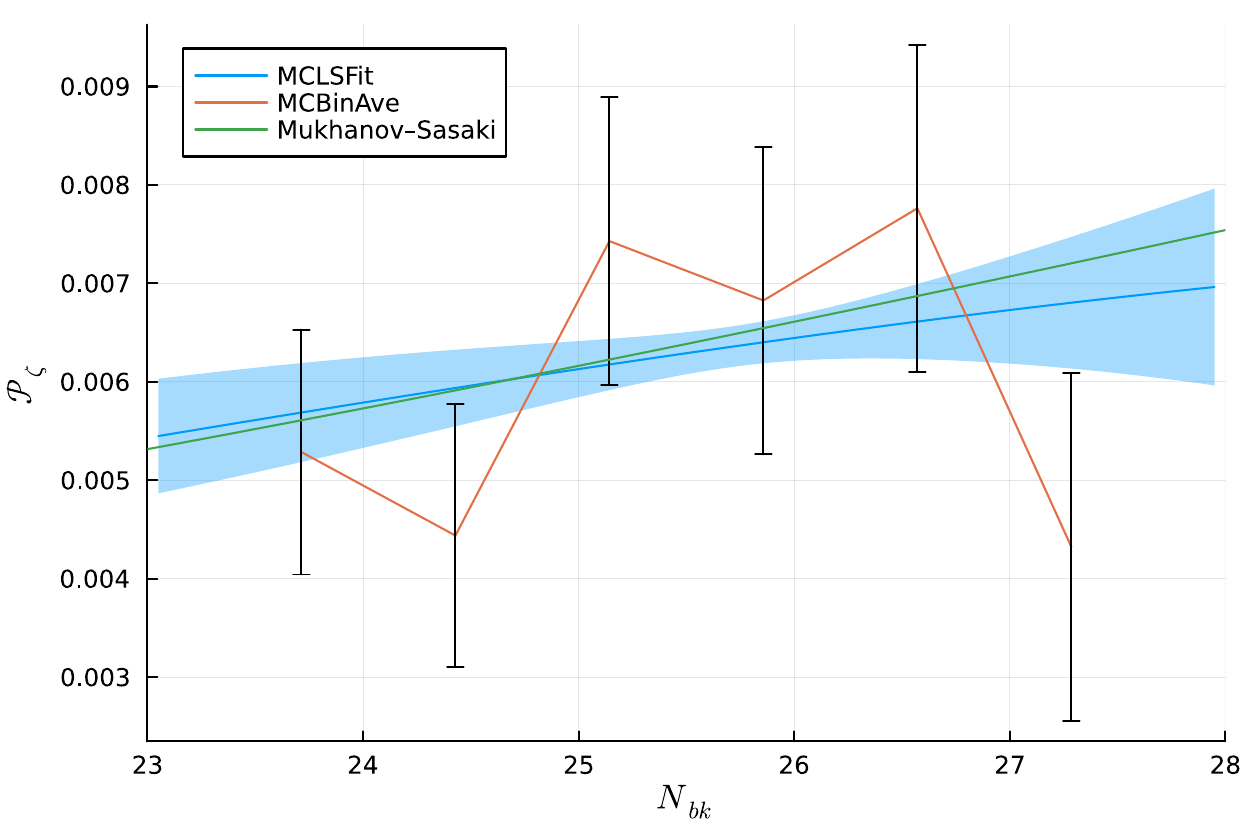}
        \end{minipage}
    \end{tabular}
    \caption{The approximating function of $F_{\left\langle\delta\mathcal{N}^2\right\rangle}$ (resp. $\mathcal{P}_\zeta$) in chaotic inflation output by MCLSFit is shown in the left panel (resp. the right panel) as a blue curve with the error shown as a blue band.
    The estimations by MCBinAve with $7$ bins are shown as red curves with the error bars. All the errors are of $1\sigma$ level. In the right panel, $\mathcal{P}_\zeta$ calculated via solving the \ac{MS} equation, is also shown in green.}
    \label{fig:chaotic}
\end{figure*}

First, we consider the single-field chaotic inflation~\cite{Linde:1983gd}, where the potential is given by
\begin{align}
V(\phi)=\frac{1}{2}m^2\phi^2,
\end{align}
with the inflaton mass $m$.
We set the model parameters as $m=0.0211$, $\phi_{\rm ini}=11$, and $\pi_{\rm ini}=-\sqrt{\frac{2}{3}}m$, and define the \ac{EOI} surface by $\epsilon_H=0.3$,\footnote{Note that $\epsilon_H=0.3$ means that inflation is still occurring. Nevertheless, we can set the \ac{EOI} surface like this because the inflatons' dynamics has been well converged to the slow-roll attractor at this point and thus $\delta N$ generated afterward is negligible. A similar discussion validates the choice of the \ac{EOI} surface in Starobinsky's linear potential model too.} where
\begin{align}
    \epsilon_H\coloneqq -\frac{\dot{H}(\boldsymbol{\varphi}, \boldsymbol{\varpi})}{H^2(\boldsymbol{\varphi}, \boldsymbol{\varpi})} = \frac{3}{2}\frac{\boldsymbol{\varpi}^2}{\frac{1}{2}\boldsymbol{\varpi}^2+V(\boldsymbol{\varphi})}.
\end{align}
This case falls into the ordinary slow-roll case, and thus the standard \ac{MS} equation-based linear perturbation analysis is applicable. 
We can compare the result of our method with the \ac{MS}-based result.
The power spectrum of the inflaton’s noise is calculated up to the next-to-leading order in the slow-roll approximation as~\cite{Mizuguchi:2024kbl}
\bme{
    \mathcal{P}_\phi(N,\varphi)=\left(\frac{H}{2\pi}\right)^2\left(\frac{\sigma\mathsf{H}}{2H}\right)^{-6\epsilon_V+2\eta_V}\left[1+\epsilon_V(10-6\gamma-12\ln2)-2\eta_V(2-\gamma-2\ln2)\right],
}
where $\gamma$ is Euler’s constant and the slow-roll parameters $\epsilon_V$ and $\eta_V$ are defined as
\begin{align}
    \epsilon_V=\frac{1}{2}\left(\frac{V^\prime(\varphi)}{V(\varphi)}\right)^2 \qc
    \eta_V=\frac{V^{\prime\prime}(\varphi)}{V(\varphi)}.
\end{align}

In our method, $N_{\rm bk}$ is sampled between  $N_{\rm bk,l}=23$ and $N_{\rm bk,u}=28$.
We take the following fitting function
\begin{align}
f(N_{\mathrm{bk}},\boldsymbol{\theta})=\exp\left(\sum_{\ell=0}^L \theta_\ell p_\ell\left(\frac{2N_{\mathrm{bk}}-N_{\rm bk,u}-N_{\rm bk,l}}{N_{\rm bk,u}-N_{\rm bk,l}}\right) \right),
\label{eq:fitFuncExp}
\end{align}
where $p_\ell$ is the $\ell$-th Legendre polynomial and $L=2$.
This choice, an exponential function with a polynomial inside, is compatible with our knowledge that in the slow-roll approximation, $\mathcal{P}_\zeta(k)$ takes a power-law form: $\mathcal{P}_\zeta(k) \simeq Ak^{\ns-1}$ with constants $A$ and $\ns$. 
Besides, this form, rather than a simple polynomial, has an advantage that the positivity of the fitted  $F_{\left<\delta\mathcal{N}^2\right>}$ is guaranteed.
To tune $\boldsymbol{\theta}$, we use \texttt{LsqFit.jl}~\cite{LsqFit}, which will also be used later.

The results of our methods MCLSFit and MCBinAve are shown in FIG.~\ref{fig:chaotic} along with $1\sigma$-level standard errors, the doubles of Eqs.~\eqref{eq:LSFitError}, \eqref{eq:LSFitErrorDeriv}, \eqref{eq:errBindelNSq}, and \eqref{eq:errBinPS}.
For $\left\langle\delta\mathcal{N}^2\right\rangle$, the results of MCLSFit and MCBinAve match well, which implies that the least squares fit is working.
Regarding the power spectrum, the MCLSFit result fits the \ac{MS}-based one, demonstrating that the proposed method as a whole is working well.
The MCBinAve result looks non-smooth due to the statistical error, but it still roughly fits the MCLSFit and \ac{MS}-based one.
Generating the samples $\{(N_{n,1}, N_{n,2})\}_n$, which is the most time-consuming part in both MCLSFit and MCBinAve, took $8.5$ minutes. 

Lastly, we remark that, although the fitting function in Eq.~\eqref{eq:fitFuncExp} reflects the prior knowledge of the power-law power spectrum, this functional form would generally be a plausible choice when no clear structure such as a peak in $\mathcal{P}_\zeta$ is implied by the result of MCBinAve.
On the other hand, if it is implied that $\mathcal{P}_\zeta$ has a peak, a functional form that naturally yields a peak in $\mathcal{P}_\zeta$ would be a better choice, as seen below.

\subsection{Starobinsky's linear-potential model \label{sec:Staro}}

\begin{figure*}
    \centering
    \begin{tabular}{c}
        \begin{minipage}{0.48\hsize}
            \centering
                \includegraphics[width=0.95\hsize]{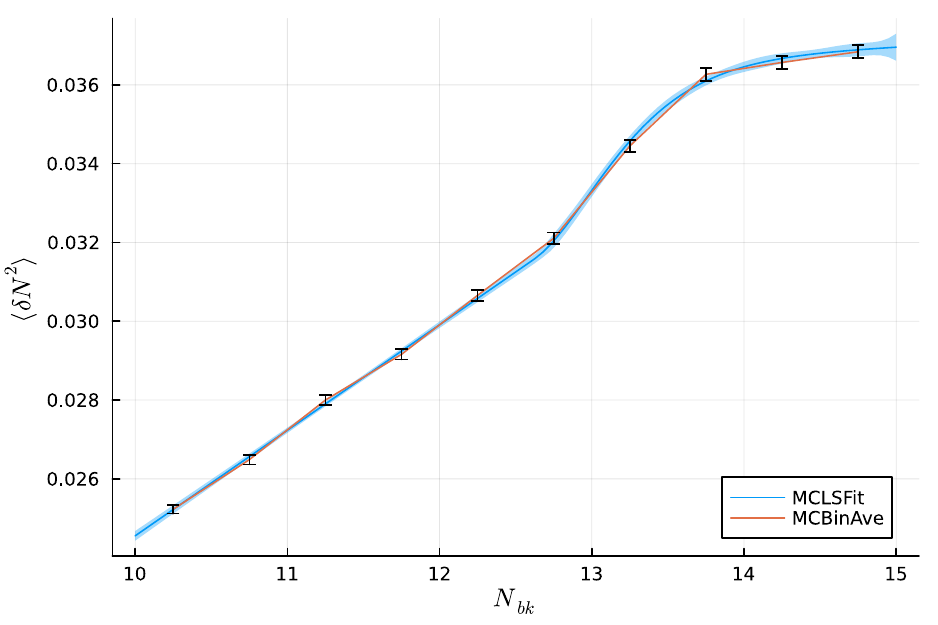}
        \end{minipage}
        \begin{minipage}{0.48\hsize}
            \centering
                \includegraphics[width=0.95\hsize]{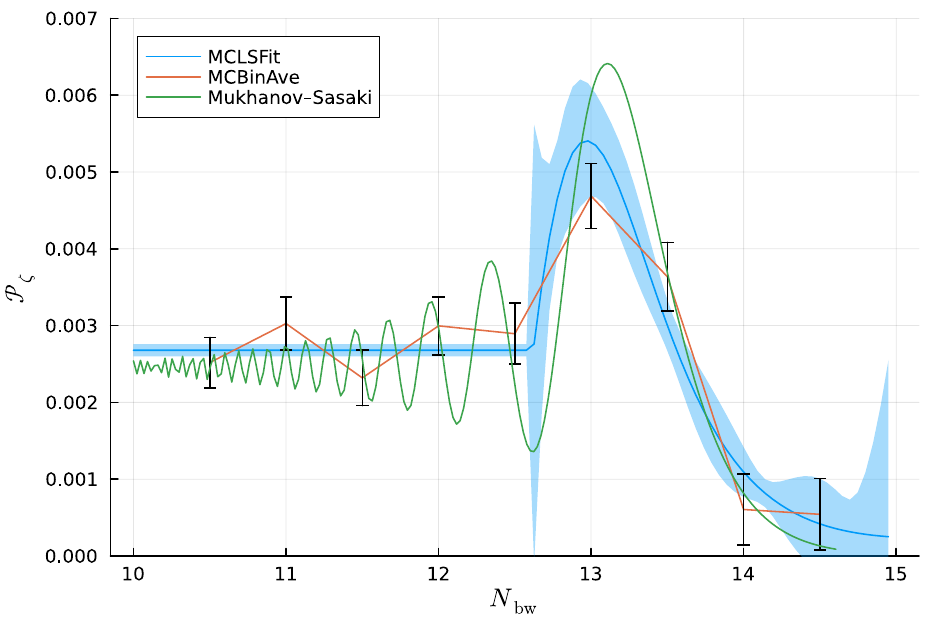}
        \end{minipage}
    \end{tabular}
    \caption{The same figure as FIG.~\ref{fig:chaotic} but for Starobinsky's linear-potential model. The number of bins in MCBinAve is now $10$.}
    \label{fig:starobinsky}
\end{figure*}

The next example is Starobinsky's linear-potential model~\cite{Starobinsky:1992ts}, a single-field model with the following potential
\begin{align}
    V(\phi)=
    \begin{dcases}
        V_0+A_+(\phi-\phi_0) & {\rm for} \ \phi\ge\phi_0, \\
        V_0+A_-(\phi-\phi_0) & {\rm for} \ \phi\le\phi_0,
    \end{dcases}
\end{align}
where $V_0$ and $A_\pm$ are positive parameters and $\phi_0$ is a kink of the potential.
We now consider the case that $A_+>A_-$ and the inflaton rolls down from $\phi_{\rm ini}>\phi_0$.
In this case, the inflaton's terminal velocity in the region $\phi>\phi_0$ is larger than that in the region $\phi<\phi_0$.
As a result, the inflaton undergoes the friction-dominated phase called \ac{USR} for a while after passing $\phi_0$, leading to the amplification of the curvature perturbation.
The inflaton's noise power spectrum $\mathcal{P}_\phi$ is well approximated by $\mathcal{P}_\phi^{1/2}=\frac{H}{2\pi}$ when $\phi>\phi_0$.
In the second phase, $\phi<\phi_0$, the following approximation formula is known in the constant-Hubble approximation~\cite{Pi:2022zxs}:
\bae{
    &\calP_{\phi}(N,\varphi,\varpi) \nonumber \\ &=\pqty{\frac{H_0}{2\pi}}^2\times\frac{1}{2 \alpha ^6 \Lambda ^2 \sigma ^6}\left[3 \left(\Lambda ^2 \left(\alpha ^4 (4 \alpha -7) \sigma ^6+\alpha ^3 (7 \alpha -16)
    \sigma ^4+(3-12 \alpha ) \sigma ^2-3\right) \right.\right. \nonumber \\
    &\qquad+\Lambda  \left(2 (5-2 \alpha ) \alpha ^4
    \sigma ^6+2 (14-5 \alpha ) \alpha ^3 \sigma ^4+6 (4 \alpha -1) \sigma ^2+6\right) \nonumber \\
    &\qquad\left.-3
    \left(\alpha ^4 \sigma ^6-(\alpha -4) \alpha ^3 \sigma ^4+(4 \alpha -1) \sigma^2+1\right)\right) \cos (2 (\alpha -1) \sigma ) \nonumber \\
    &\qquad+\left(\sigma ^2+1\right) \left(-18
    \Lambda  \left(\alpha ^2 \sigma ^2+1\right)^2+9 \left(\alpha ^2 \sigma^2+1\right)^2+\Lambda ^2 \left(2 \alpha ^6 \sigma ^6+9 \alpha ^4 \sigma ^4+18 \alpha ^2
    \sigma ^2+9\right)\right) \nonumber \\
    &\qquad+6 \sigma  \left(\alpha ^5 (\Lambda -1) \Lambda  \sigma ^4
    \left(\sigma ^2-1\right)+\alpha ^4 \left(7 \Lambda ^2-10 \Lambda +3\right) \sigma^4-\alpha ^3 \left(4 \Lambda ^2-7 \Lambda +3\right) \sigma ^2 \left(\sigma ^2-1\right) \right. \nonumber \\
    &\qquad\left.\left.-3
    \alpha  (\Lambda -1)^2 \left(\sigma ^2-1\right)-3 (\Lambda -1)^2\right) \sin (2 \sigma -2
    \alpha  \sigma )\right],
}
Here, $H_0$ is the Hubble parameter at the initial time, $\Lambda\coloneqq A_+/A_-$, and $\alpha\coloneqq\exp(N-\mathcal{N}_0)$ with the transition time $\mathcal{N}_0$ defined as $\varphi(\mathcal{N}_0)=\phi_0$, which is also a random variable different among paths.
Now, the model parameters are set same as~\cite{Mizuguchi:2024kbl}: $V_0=3H_0^2$, $H_0=10^{-5}$, $A_+=\sqrt{\frac{9H_0^6}{4\pi^2 \times 8.5\times10^{-10}}}$, $A_-=A_+/1700$, $\varphi_{\rm ini}=1.93\times10^{-2}$, $\varpi_{\rm ini}=-5.45\times10^{-7}$.
The \ac{EOI} is defined as the time when $\phi$ reaches $\phi_{\rm end}=-0.0187$.

The setting of our method is as follows.
Only in this test case, as the scheme for discretizing Eq.~\eqref{eq:SDE}, we use not the \ac{EM} scheme but a higher-degree one \texttt{SRIW2}~\cite{Rossler2010} in \texttt{DifferentialEquations.jl} with a finer time step size $\Delta N=0.001$, since solving Eq.~\eqref{eq:SDE} in the \ac{USR} requires the higher accuracy.
$n_{\rm path}$ is also set to a larger value, $10^6$.
We set $N_{\rm bk,l}=10$ and $N_{\rm bk,u}=15$.
The fitting function is taken as the following one, consisting of two parts:
\begin{align}
    f(N_{\mathrm{bk}},\boldsymbol{\theta})=  
    \begin{dcases}
         f^{\rm log}_{L,N_{\rm bk,u},N_{\rm bk,l}}(N_{\mathrm{bk}},\boldsymbol{\theta})& \text{for } N_{\mathrm{bk}}>\theta_{-2}, \\
        aN_{\mathrm{bk}}+b & \text{otherwise}, 
    \end{dcases}
    \label{eq:fStaro}
\end{align}
where
\bae{ 
    f^{\rm log}_{L,N_{\rm bk,u},N_{\rm bk,l}}(N_{\mathrm{bk}},\boldsymbol{\theta})
    \coloneqq\frac{\theta_{-1}}{1+\exp\left(\sum_{\ell=0}^L \theta_\ell p_\ell\left(\frac{2N_{\mathrm{bk}}-N_{\rm bk,u}-N_{\rm bk,l}}{N_{\rm bk,u}-N_{\rm bk,l}}\right) \right)}.
}
This function has $L+3$ tunable parameters $\boldsymbol{\theta}=(\theta_{-2},\theta_{-1},...,\theta_{L})$, the connection point $\theta_{-2}$ and the parameters $\theta_{-1},...,\theta_{L}$ that determines the right part. 
The parameters $a$ and $b$ for the left part are automatically determined so that $f$ and $\partial_{N_{\rm bk}}f$, which correspond to $\left\langle \delta \mathcal{N}^2\right\rangle$ and $\mathcal{P}_\zeta$ respectively, are continuous at $N_{\rm bk}=\theta_{-2}$.
We now set $L=4$.
We will discuss this choice of the fitting function later.

The result is shown in FIG.~\ref{fig:starobinsky}.
The \ac{MS}-based calculation is applicable in this case~\cite{Mizuguchi:2024kbl} and its result is also shown in the figure.
The \ac{USR} yields a characteristic shape of $\mathcal{P}_\zeta$: the largest peak around the point corresponding to the inflaton's passing the potential kink and the oscillation on a flat baseline after that.
MCBinAve roughly reproduces this shape, and so does MCLSFit, although the oscillation on the baseline is not captured with the current fitting function.
The sample generation took $17.1$ hours in this case, because of the small $\Delta N$ and large $n_{\rm path}$.

We finally remark on the choice of the fitting function.
We chose this by taking into account the shape of the power spectrum, specifically the large peak on the right and the otherwise flat line with oscillations on the left.
In fact, the right part of $f$ in Eq.~\eqref{eq:fStaro} is based on the logistic function $1/(1+e^{-x})$, which has a sigmoid shape, and thus its single-peaked derivative naturally fits the largest peak of $\mathcal{P}_\zeta$.
The left part of Eq.~\eqref{eq:fStaro} is a linear function, whose derivative is constant.
With this choice, we have obtained a function fitting the true power spectrum to some extent, as shown in FIG.~\ref{fig:starobinsky}.
It may seem like a kind of cheat that we choose the fitting function knowing the true function.
Although this is true, it is common and desirable to reflect prior knowledge and/or rough estimates on the true function in the machine learning context.
In the current case, even if we do not know the true function, the MCBinAve estimate roughly traces its shape, suggesting a fitting function shape similar to Eq.~\eqref{eq:fStaro}.

\subsection{Flat quantum well \label{sec:well}}

\begin{figure*}
    \centering
    \begin{tabular}{c}
        \begin{minipage}{0.48\hsize}
            \centering
                \includegraphics[width=0.95\hsize]{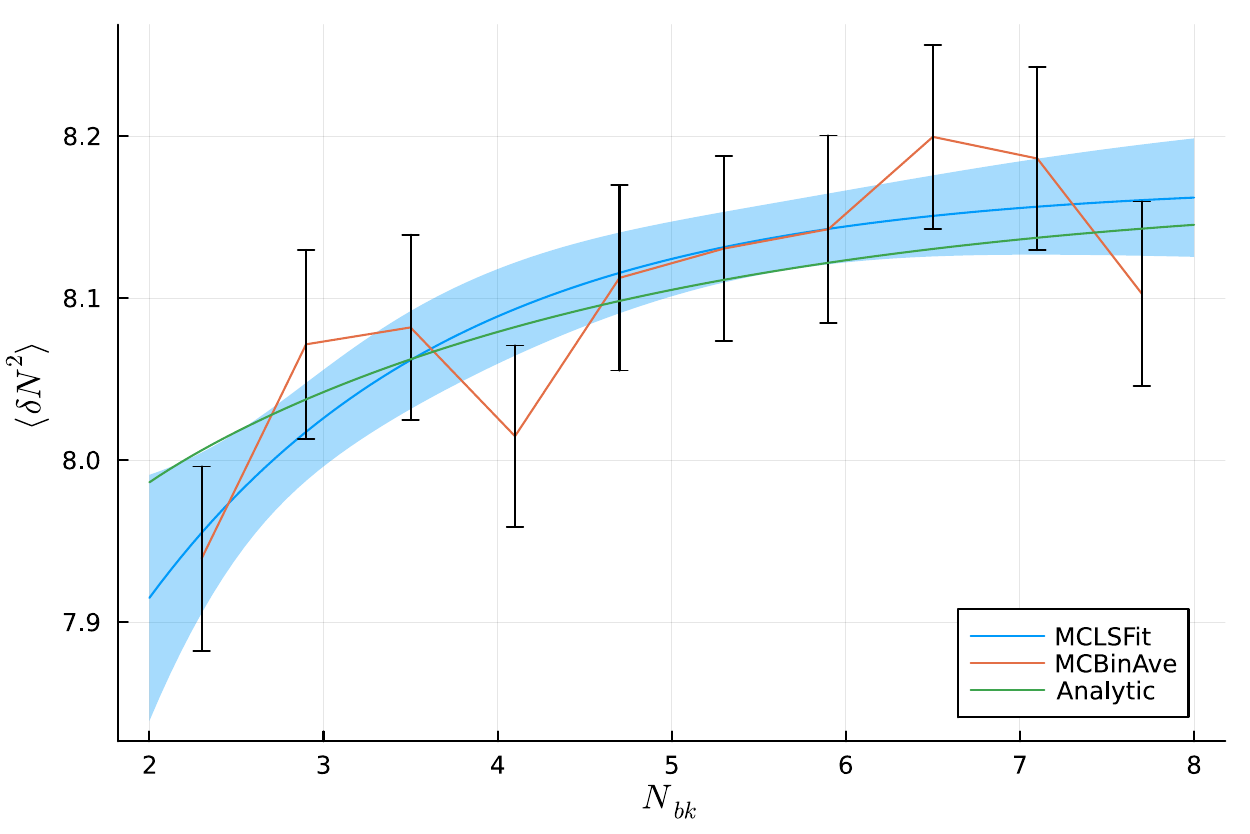}
        \end{minipage}
        \begin{minipage}{0.48\hsize}
            \centering
                \includegraphics[width=0.95\hsize]{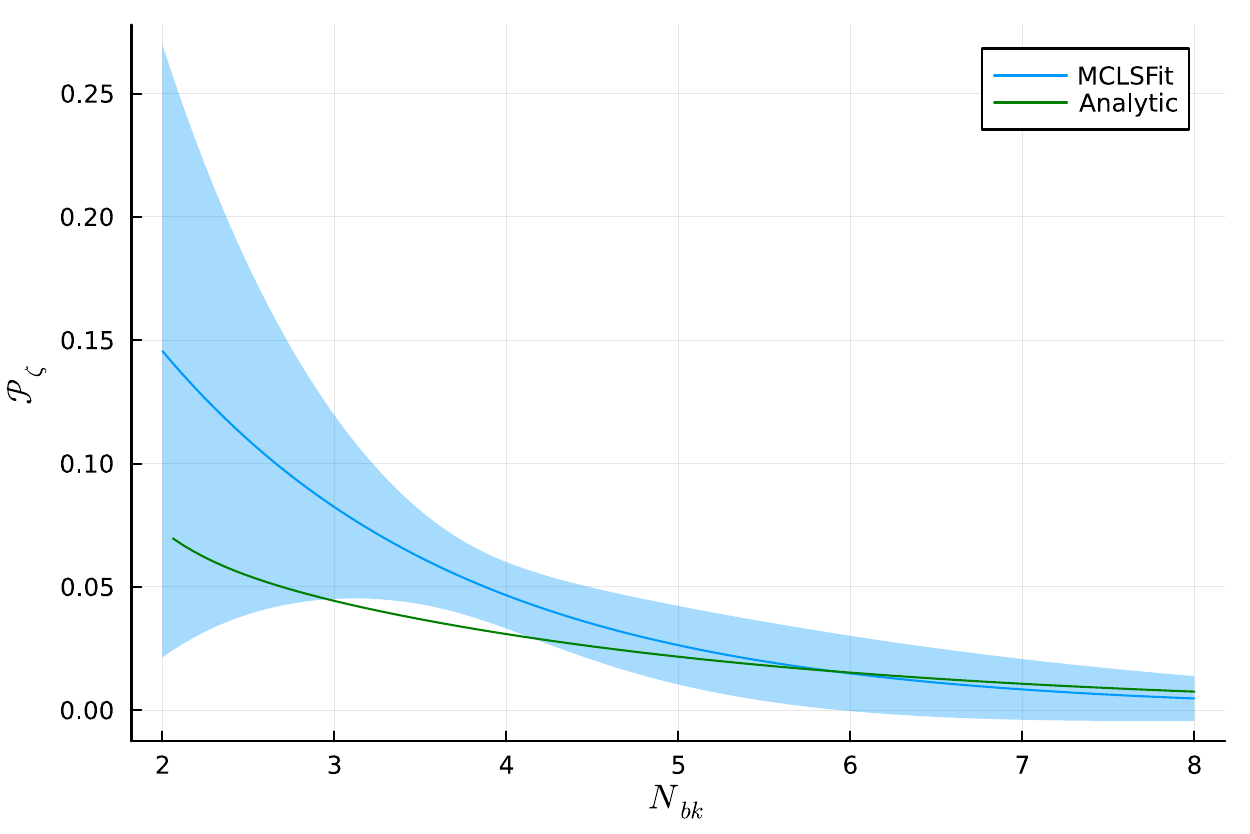}
        \end{minipage}
    \end{tabular}
    \caption{The same figure as FIG.~\ref{fig:chaotic} but for the flat quantum well model. The number of bins in MCBinAve is now $10$. The analytical formulae~\eqref{eq: AV variance} and \eqref{eq: AV power}
    are shown in green. In the right panel, the MCBinAve result is not shown because the statistical error bar is so large that plotting the result is not illustrative.}
    \label{fig:flat}
\end{figure*}

The above two examples are the cases where the quantum diffusion does not dominantly determine the inflaton's dynamics: the stochastic formalism is not necessarily required, and in fact, the MS approach works.
As a third example, let us consider the flat quantum well model in \ac{AV20}~\cite{Ando:2020fjm}.
This is a single-field model, in which the potential $V(\phi)$ is completely flat, namely $V(\phi)=24\pi^2v_{\rm well}={\rm const}$ in the interval $\bqty{\phi_\uw,\phi_\uc}$.
Thus, in this interval, there is no potential force on the inflaton, and its dynamics is completely dominated by the quantum diffusion.
At the ends of the interval, the potential is continuously connected to the non-flat parts: for $\phi<\phi_\uw$, $V(\phi)$ is not flat but so shallow that $\phi$ shows the slow-roll dynamics, and $V(\phi)$ is so steep for $\phi>\phi_\uc$ that $\phi_\uc$ is the \ac{EOI} point.\footnote{Although \ac{AV20}~\cite{Ando:2020fjm} considered a more general case that $\phi$ can slow-roll also for $\phi>\phi_\uc$, we now set $\phi_\uc$ to the \ac{EOI} point for simplicity as it merely shifts the scale of interest.}
In this setting, the inflaton's dynamics is described by the following one-dimensional Langevin equation~\cite{Ando:2020fjm}
\begin{align}
    \frac{{\rm d}x(N)}{{\rm d} N} = \frac{\sqrt{2}}{\mu}\xi(N),
    \label{eq:SDEFlat}
\end{align}
where $x(N)\coloneqq(\varphi(N)-\phi_\uw)/(\phi_\uc-\phi_\uw)$ and $\mu=(\phi_\uc-\phi_\uw)/\sqrt{v_\well}$.
$x(N)$ has a reflective boundary at $x=0$ and an absorbing boundary at $x=1$, which means that $x(N)$ stops when it reaches $x=1$, corresponding the \ac{EOI}.
Further, it is equivalent to the absolute value of a non-reflective process $x(N)$ that obeys Eq.~\eqref{eq:SDEFlat} and stops when $|x(N)|=1$.
After all, to sample a total e-fold number of the inflaton's path starting from $\phi_{\rm ini}$, we may simulate $x(N)$ by Eq.~\eqref{eq:SDEFlat} with the initial value $x_\ini=(\phi_\ini-\phi_\uw)/(\phi_\uc-\phi_\uw)$ and take the time when $x(N)$ first reaches $1$ or $-1$.

Practically, a naive simulation of Eq. \eqref{eq:SDEFlat} with this \ac{EOI} condition by the \ac{EM} method may not lead to accurate sampling of the first passage time in the currently considered diffusion-dominant system.
For a simple problem, the first passage time of a one-dimensional Brownian motion, as we are now considering, we can use improved sampling methods developed in the field of probability theory.
Specifically, we adopt the boundary shifting method proposed in Ref.~\cite{Broadie1997} for a problem in mathematical finance: we reset the levels at which $x(N)$ stops as
\begin{align}
    \pm \left(1-\frac{0.5826\sqrt{2\Delta N}}{\mu}\right).
\end{align}
For a more detailed explanation on this method, see Appendix~\ref{sec:BoundShift}.

We compare our results in MCLSFit and MCBinAve with \ac{AV20}'s analytic formulae~\cite{Ando:2020fjm}:
\bae{
    \left\langle\delta\mathcal{N}^2\right\rangle(N_{\rm bk})=\frac{\mu^4}{6}- \frac{8\mu^4}{3\pi^2}\sum_{n=0}^\infty \frac{e^{-\left(n+\frac{1}{2}\right)^2\frac{\pi^2}{\mu^2}N_{\rm bk}}}{\left(n+\frac{1}{2}\right)^2} 
    - \frac{8\mu^4}{\pi^5}\sum_{n=0}^\infty \frac{e^{-\left(n+\frac{1}{2}\right)^2\frac{\pi^2}{\mu^2}N_{\rm bk}}}{\left(n+\frac{1}{2}\right)^4}\left[5\frac{(-1)^n}{n+\frac{1}{2}}-4\pi\right].
    \label{eq: AV variance}
}
and
\bae{\label{eq: AV power}
    \mathcal{P}_\zeta(N_{\rm bk})=\frac{8\mu^2}{3}\sum_{n=0}^\infty e^{-\left(n+\frac{1}{2}\right)^2\frac{\pi^2}{\mu^2}N_{\rm bk}} 
    + \frac{8\mu^2}{\pi^3}\sum_{n=0}^\infty \frac{e^{-\left(n+\frac{1}{2}\right)^2\frac{\pi^2}{\mu^2}N_{\rm bk}}}{\left(n+\frac{1}{2}\right)^2}\left[5\frac{(-1)^n}{n+\frac{1}{2}}-4\pi\right], 
}
We truncate the infinite sum at $n=10$.
In particular, we would like to see the perturbation for $N_\bk$ larger than the expected e-folds during the quantum well phase, $\expval{\calN(x=0)}=\mu^2/2$, as a very characteristic, stochastic effect reported by \ac{AV20}: it corresponds to larger scales than the quantum-well scale but the stochastic effect predicts such a ``leakage" of small-scale perturbations to large-scale ones.

The results of our methods for $\mu=\sqrt{7}$, corresponding to, $\expval{\calN(x=0)}=3.5$, are shown in FIG.~\ref{fig:flat}, along with these analytical solutions. 
We set $N_{\rm bk,l}=3$ and $N_{\rm bk,u}=8$.
In path generation, we took a finer time step size $\Delta N=0.001$ and a larger sample size $n_{\rm path}=10^6$, in order to accurately reproduce the curve of $\left\langle\delta\mathcal{N}^2\right\rangle$, which shows a larger value but a smaller relative variation than in other test cases, and to subsequently obtain $\mathcal{P}_\zeta$.
The initial point of paths should be set well before the quantum well phase to see the ``leakage" effect, but we neglect the perturbation during the slow-roll phase, and hence we can set the initial point to $x_{\rm ini}=0$, which corresponds to the reflective boundary of the well. 
We found that, for a part of the paths sampled in step~3 in ALGORITHM~\ref{alg:main}, say the $n$-th one, its total e-fold $\calN_n$ was smaller than $N_{{\rm bk},n}$, which made how to set $\mathbf{X}_{\mathrm{bk},n}=x_{\mathrm{bk},n}$, the value of $x$ $N_{{\rm bk},n}$ e-folds before the \ac{EOI}, unclear.
In such a case, we can again set $x_{\mathrm{bk},n}=0$ by neglecting the perturbation during the slow-roll phase.
For fitting, we used the following function
\begin{align}
f(N_{\mathrm{bk}},\boldsymbol{\theta})=\theta_{1}+\theta_2\exp\left(\theta_3 N_{\mathrm{bk}}\right)
\label{eq:fitFlat}
\end{align}
with three parameters $\boldsymbol{\theta}=(\theta_1,\theta_2,\theta_3)$.
We took this function shape considering the analytical formula of $\left\langle\delta\mathcal{N}^2\right\rangle$, for which taking only the leading terms yields
\begin{align}
    \left\langle\delta\mathcal{N}^2\right\rangle(N_{\rm bk})\simeq\frac{\mu^4}{6}- \frac{32\mu^4}{3\pi^2}\left(1-\frac{48}{\pi^2}+\frac{120}{\pi^3}\right)e^{-\frac{\pi^2}{4\mu^2}N_{\rm bk}}.
    \label{eq:delN2FlatApp}
\end{align}
Again, this choice is also suggested by the MCBinAve result of $\left\langle\delta\mathcal{N}^2\right\rangle$, which gradually approaches a horizontal line from below.
FIG.~\ref{fig:flat} shows that although the statistical error is rather large, our estimates of $\left\langle\delta\mathcal{N}^2\right\rangle$ and $\mathcal{P}_\zeta$ closely coincide with the true ones.
The ``leakage" effect for $N_\bk>\expval{\calN(x=0)}=3.5$ can also be seen.
Sample generation took $8.5$ hours in this case.

Let us comment on the formulation dependence of the ``leakage" effect. This effect is because the probability of $\calN_n>N_{\bk,n}$ is not negligible even for $N_{\bk,n}>\expval{\calN(x=0)}$ in such a diffusion-dominant system, and the quantum-well fluctuation can account for a part of the large-scale perturbations.
This does not happen in \ac{FKTT} or \ac{AV24}, as they choose $\bfX$ based on the average e-folds.
In this paper, we do not discuss which formulation is correct, but we merely reproduce \ac{AV20}'s result in our method based on Monte Carlo simulation and least squares fitting.

\subsection{Hybrid inflation}

\begin{figure*}
    \centering
    \begin{tabular}{c}
        \begin{minipage}{0.48\hsize}
            \centering
                \includegraphics[width=0.95\hsize]{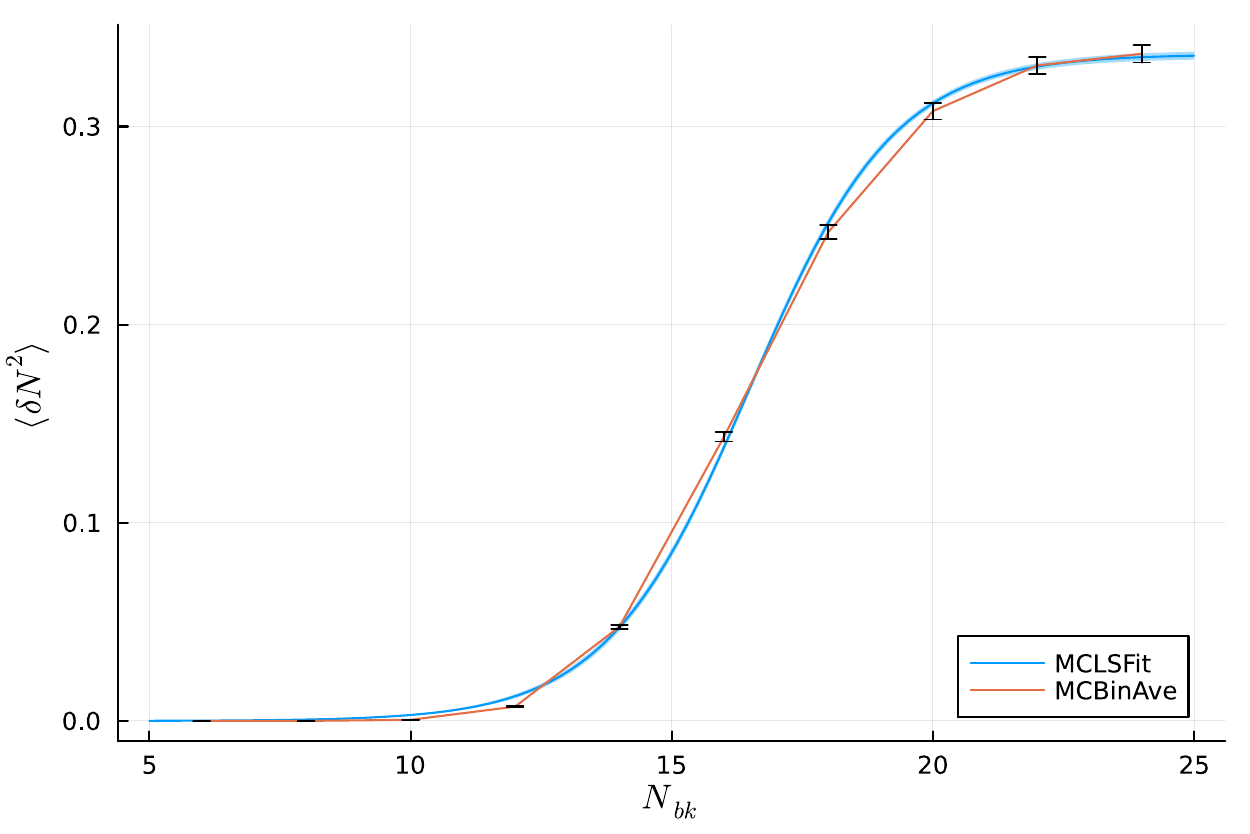}
        \end{minipage}
        \begin{minipage}{0.48\hsize}
            \centering
                \includegraphics[width=0.95\hsize]{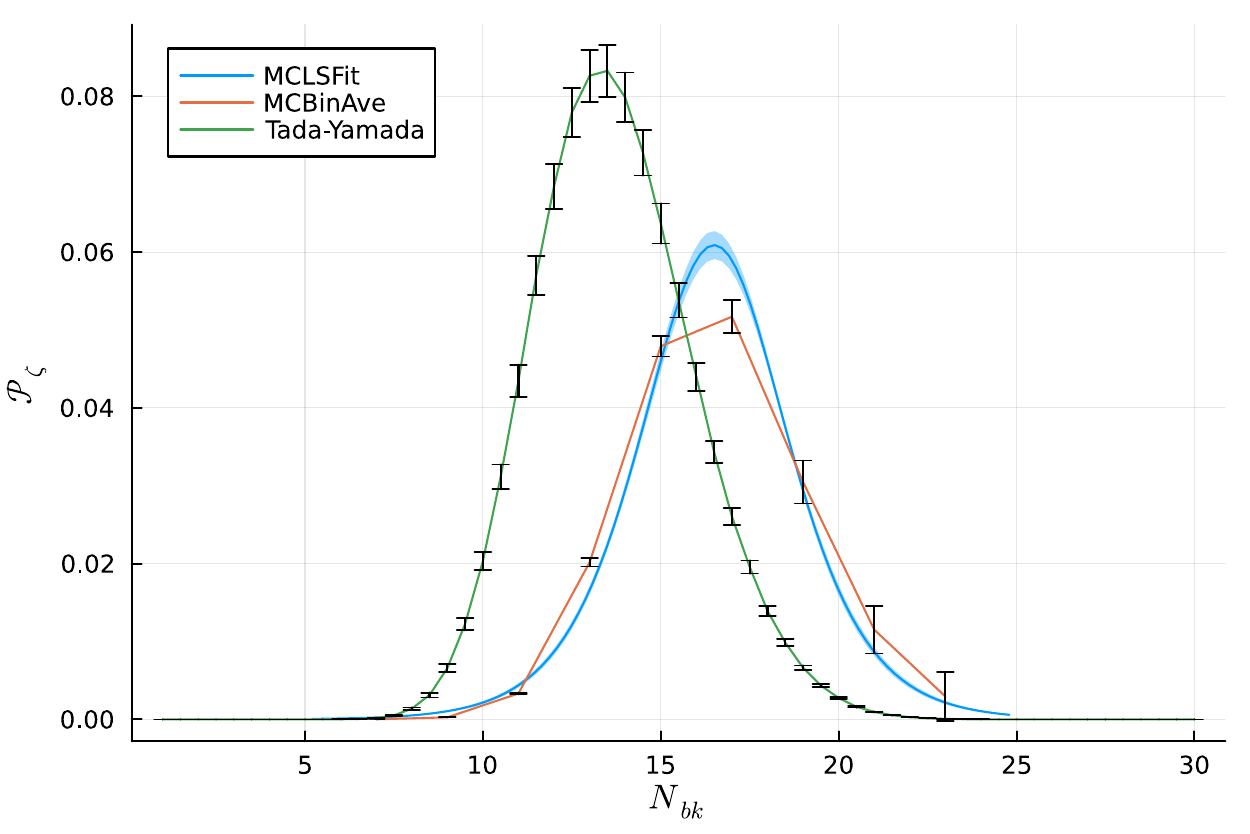}
        \end{minipage}
    \end{tabular}
    \caption{The same figure as FIG.~\ref{fig:chaotic} but for hybrid inflation. The number of bins in MCBinAve is $10$. In the right panel, the power spectrum calculated by the method in Ref.~\cite{Tada:2023fvd} is shown in green.}
    \label{fig:hybrid}
\end{figure*}

The last example is hybrid inflation~\cite{Linde:1993cn} with the potential
\begin{align}
    V(\boldsymbol{\phi})= \Lambda^4\left[\left(1-\frac{\psi^2}{M^2}\right)^2+2\frac{\phi^2\psi^2}{\phi_\uc^2M^2}+\frac{\phi-\phi_\uc}{\mu_1}-\frac{(\phi-\phi_\uc)^2}{\mu_2^2}\right],
\end{align}
where $\Lambda,M,\mu,\phi_\uc>0$ are the parameters.
This model involves two fields $\boldsymbol{\phi}=(\phi,\psi)$, which are usually called the inflaton and the waterfall field.
Starting with the initial values $\phi_{\rm ini}>\phi_\uc$ and $\psi_{\rm ini}=0$, $\phi$ rolls down with $\psi$ kept around $0$, the minimum point in the $\psi$ direction.
After it reaches $\phi_\uc$, $\psi=0$ becomes the local maximum point, which is called the waterfall transition, and $\boldsymbol{\phi}$ rolls down to either of the global minima $(0,\pm M)$.
Depending on the parameters, the potential around the inflection point $\boldsymbol{\phi}=(\phi_\uc,0)$ is very flat, and thus $\boldsymbol{\phi}$ experiences the diffusion-dominated phase before inflation ends with $\epsilon_H=1$.
This implies that the curvature perturbation of a scale that exits the Hubble horizon around the waterfall transition may be greatly enhanced, as explored in, e.g., Refs.~\cite{Garcia-Bellido:1996mdl,Lyth:2010zq,Bugaev:2011qt,Bugaev:2011wy,Lyth:2012yp,Guth:2012we,Halpern:2014mca,Fujita:2014tja,Clesse:2015wea,Kawasaki:2015ppx,Tada:2023pue,Tada:2023fvd}.

Now, as in Ref.~\cite{Tada:2024ckk}, we set the model parameters as
\bege{
    M=\frac{\phi_\uc}{\sqrt{2}}=10^{16}\,\si{GeV} \qc \mu_1=\frac{100}{M^2\phi_\uc} \qc \mu_2=10 \qc 
    \Lambda=5.4\times10^{15}\,\si{GeV} \times M\phi_\uc^{1/2}, 
}
and the initial values
\begin{align}
    (\boldsymbol{\phi}_{\rm ini},\boldsymbol{\pi}_{\rm ini})=(\phi_{\rm ini},\psi_{\ini},0,0), 
\end{align}
where
\bae{
    \phi_{\rm ini}=\phi_\uc+\frac{15}{\mu_1} \qc \psi_{\rm ini}=\sqrt{\frac{5\Lambda^4}{24\sqrt{2\pi^3}}}.
}
For each field, the noise power spectrum is set as $\mathcal{P}_\phi=\mathcal{P}_\psi=\frac{H}{2\pi}$.
The EOI is set to the moment when $\partial_{\psi}^2V/V=-2$.
For the setting of our method, we set $N_{\rm bk,l}=5$ and $N_{\rm bk,u}=25$ and take the fitting function $f(N_{\mathrm{bk}},\boldsymbol{\theta})=f^{\rm log}_{L,N_{\rm bk,u},N_{\rm bk,l}}(N_{\mathrm{bk}},\boldsymbol{\theta})$, a logistic function with a polynomial inside, which was also used in Sec.~\ref{sec:Staro}.
We chose this function because of the knowledge that $\mathcal{P}_\zeta$ has a single large peak; the derivative of a logistic function has such a shape.
The single-peak shape of $\mathcal{P}_\zeta$ is also implied by MCBinAve, as seen shortly below, and in a general case where MCBinAve implies such a shape, the logistic function-based $f(N_{\mathrm{bk}},\boldsymbol{\theta})$ would be a plausible choice.
We now set $L=2$.

We show the result of our method in FIG.~\ref{fig:hybrid}.
Now, the \ac{MS}-based result is not accompanied because the perturbed approach breaks down in this model.
MCLSFit produces $\tilde{\mathcal{P}}_\zeta$ having a peak with small statistical errors.
The result of MCBinAve almost fits this, with the peak height somewhat suppressed due to its averaging nature.

For the sake of comparison, we also plot $\mathcal{P}_\zeta$ calculated by the method proposed by \ac{TY}~\cite{Tada:2023fvd}, based on the adjoint Fokker--Planck \ac{PDE} described in Sec.~\ref{sec: Animali}.
Except for several tens of percent differences in the position and the height of the peak, their qualitative features agree with each other.
Quantitative differences may arise from the fact that \ac{TY} took the slow-roll approximation to neglect the velocities $\bm{\varpi}$.
In this test case, our code took $9.8$ minutes for sample generation, while \ac{TY}'s programme takes around $40$ minutes in addition to $6$ minutes to solve the \acp{PDE} (by M1 Max MacBook Pro), though \ac{TY}'s sample generation would also be improved by utilizing a public library.

\section{Summary \label{sec:sum}}

In this paper, we proposed novel algorithms to calculate the inflationary curvature perturbation power spectrum $\mathcal{P}_\zeta$ in the stochastic-$\delta \calN$ formalism.
To deal with multi-field inflation models, some kind of numerical approach is desired, but existing Monte Carlo-based methods are computationally demanding due to their nested structure.
Our proposal makes improvements twofold.
First, we avoid nested Monte Carlo simulation by taking an estimator \eqref{eq:delNSqEstimator} of $\left\langle \delta \mathcal{N}_{\mathbf{X}}^2 \right\rangle$, which can be calculated by generating just two branches from each path.
This largely reduces the number of paths generated in the simulation.
Second, we incorporate least squares curve fitting into this Monte Carlo-based method: we perform fitting of a certain parametric function to data sampled in Monte Carlo path generation.
Then, our method, MCLSFit, does not need to estimate $\mathcal{P}_\zeta$ at a finite number $n_{\rm PS}$ of scales, and its computational cost is not proportional to $n_{\rm PS}$.
Rather, our method produces an approximating function of $\mathcal{P}_\zeta$ over a range of scales with a reduced computational cost.
We also conduct numerical demonstrations of our method in some inflation models.
The results imply that, although the choice of fitting function can be an issue, especially when $\mathcal{P}_\zeta$ has a complicated shape, our method works in various cases, including hybrid inflation.
We also present a supporting method, MCBinAve, based on binning and averaging the samples of the estimator \eqref{eq:delNSqEstimator}, which can be used to check the fitting accuracy and get suggestions for the form of the fitting function.
For example, if it is implied that $\mathcal{P}_\zeta$ has a single peak, the logistic function with a polynomial inside may be a good choice, and if no clear structure is implied, the exponential with a polynomial inside is a simple choice.

Although we take only the power spectrum as a target in this paper, it would be interesting to extend our method to calculating higher-order statistics.
For example, since
\begin{align}
    \left\langle \delta \mathcal{N}_{\mathbf{X}}^3 \right\rangle= \left\langle\frac{1}{6}\left(2\mathcal{N}_{\mathbf{X}}^{(1)}-\mathcal{N}_{\mathbf{X}}^{(2)}-\mathcal{N}_{\mathbf{X}}^{(3)}\right)^3\right\rangle
\end{align}
holds, where $\mathcal{N}_{\mathbf{X}}^{(1,2,3)}$ are e-folds of three independent paths starting from the point $\mathbf{X}$, we have an estimator of $\left\langle \delta \mathcal{N}_{\mathbf{X}}^3\right\rangle$ that can be calculated via generating just three branches.
If we could find some formula like Eq.~\eqref{eq:PSAndo} that connects $\left\langle \delta \mathcal{N}_{\mathbf{X}}^3 \right\rangle$ to the bispectrum of the curvature perturbation, we could build a similar method for the bispectrum.
We will consider such a possibility in future work.

\section*{Acknowledgements}

KM is supported by MEXT Quantum Leap Flagship Program (MEXT Q-LEAP) Grant no. JPMXS0120319794, and JST COI-NEXT Program Grant No. JPMJPF2014.
YT is supported by JSPS KAKENHI Grant
No.~JP24K07047.

\appendix

\section{Boundary shifting \label{sec:BoundShift}}

The task considered in Sec.~\ref{sec:well} is a kind of first-passage problem for a Brownian motion.
Let us consider a random process $X(t)$ obeying the following Langevin equation
\begin{equation}
    \dv{X(t)}{t} = \sigma \xi(t)
\end{equation}
with a constant $\sigma>0$ and the initial value $X(0)=0$, or, equivalently, the following stochastic differential equation
\begin{equation}
    \dd{X(t)} = \sigma \dd{W(t)},
\end{equation}
where $W(t)$ is a Brownian motion.
Then, we consider the time $\tau$ when $X$ first exceeds the value $B$,
\begin{align}
    \tau_B \coloneq \inf \left\{t>0 \mid X(t)\ge B \right\},
\end{align}
where $B$ is assumed to be positive without loss of generality.
To analyze the distribution of $\tau$, the approach taken in Sec.~\ref{sec:well} is the Monte Carlo-based one: we generate paths of $X$ and take
\begin{align}
    \tilde{\tau}_B = \inf \left\{t_j \mid X(t_j)\ge B \right\},
\end{align}
as an approximation of $\tau_B$.
Here, although $X$ is originally continuous in time, generated paths are inevitably discrete: we assume that each of them consists of $X(t_1),\dots,X(t_n)$, where $t_i \coloneq i\Delta t$ is the $i$-th time grid point with an interval $\Delta t$.
Then, this causes the issue of discrete observation.
For a path, if all the $X(t_1),\dots,X(t_n)$ are below $B$, we consider that this path did not exceed $B$.
However, such a path might actually exceed $B$ at some time between $t_i$ and $t_{i+1}$.
We are neglecting such an event, which leads to underestimating the probability of $X$ reaching $B$ and overestimating $\tau_B$: according to Ref.~\cite{Broadie1997},
the difference between the distributions of $\tau_B$ and $\tilde{\tau}_B$ scales as
\begin{align}
    \left|{\rm Pr}\left(\tilde{\tau}_B \le t_i\right) - {\rm Pr}\left(\tau_B \le t_i\right)\right| = O(\sqrt{\Delta t}).
\end{align}

Coping with the difference between the discrete first-passage problem and the continuous one has been studied in the field of probability theory, especially mathematical finance.
A simple way is shifting the boundary, which was proposed in Ref.~\cite{Broadie1997} based on Ref.~\cite{siegmund1982brownian}.
Because the first-passage probability is higher for a lower $B$, lowering $B$ pushes the probability underestimated by the discrete observation to the true value.
According to Ref.~\cite{Broadie1997}, by resetting $B$ to 
\begin{align}   
    \tilde{B}\coloneq B-\beta\sigma\sqrt{\Delta t} \qc 
    \beta\coloneq -\frac{\zeta\left(\frac{1}{2}\right)}{\sqrt{2\pi}}\simeq0.5826,    
\end{align}
where $\zeta$ is the Riemann zeta function, the first-passage probability with discrete observation is approximately modified to the continuous one:
\begin{align}
    \left|{\rm Pr}\left(\tilde{\tau}_B \le t_i\right) - {\rm Pr}\left(\tau_B \le t_i\right)\right| = o(\sqrt{\Delta t}).
\end{align}

\bibliographystyle{JHEP}
\bibliography{reference}

\end{document}